\documentclass[fleqn,usenatbib]{mnras}
\usepackage[utf8]{inputenc}

\usepackage{graphicx}	%
\usepackage{amsmath}	%
\usepackage{amssymb}
\usepackage[dvipsnames]{xcolor}
\usepackage{lscape}
\usepackage{booktabs}
\usepackage[normalem]{ulem}
\usepackage{fixme}

\fxsetup{
  status=draft,
  layout=inline,nomargin,nofootnote,
  theme=colorsig,
}
\FXRegisterAuthor{cc}{acc}{CC} %

\FXRegisterAuthor{rh}{arh}{RH} %
\FXRegisterAuthor{mr}{amr}{MR} %
\begingroup
\catcode`\_=\active
\gdef_#1{\ensuremath{\sb{\mathrm{#1}}}}
\endgroup
\mathcode`\_=\string"8000
\catcode`\_=12

\newcommand{\ramses}{\textsc{RAMSES}}                %

\newcommand{\msun}{\ensuremath{{\rm M}_{\odot}}}
\newcommand{\pcc}{\ensuremath{{\rm cm}^{-3}}}	%

\newcommand{\hide}[1]{}

\title[Star Formation in Magnetized GMCs]{Bimodal Star Formation in Simulations of Strongly Magnetized Giant Molecular Clouds}
\author[]{Ronan Hix$^{1}$\thanks{E-mail: rhix@terpmail.umd.edu}, Chong-Chong He$^{1}$\thanks{E-mail: che1234@umd.edu}, Massimo Ricotti$^{1}$\thanks{E-mail: ricotti@umd.edu}
\\
$^{1}$Department of Astronomy, University of Maryland, College Park, MD 20742, USA
}
\date{December 2022}

\begin{document}
\label{firstpage}
\pagerange{\pageref{firstpage}--\pageref{lastpage}}
\maketitle

\begin{abstract}
We present the results of a set of radiation magnetohydrodynamic simulations of turbulent molecular clouds in which we vary the initial strength of the magnetic field within a range ($1 \lesssim \mu \lesssim 5$) consistent with observations of local giant molecular clouds (GMCs).
We find that as we increase the strength of the magnetic field, star formation transitions from unimodal (the baseline case, $\mu=5$, with a single burst of star formation and Salpeter IMF) to bimodal. This effect is clearest in the most strongly magnetized GMC ($\mu=1$): a first burst of star formation with duration, intensity and IMF comparable to the baseline case, is followed by a second star formation episode in which only low-mass stars are formed. Overall, due to the second burst of star formation, the strongly magnetized case results in a longer star formation period and higher efficiency of star formation.
The second burst is produced by gas that is not expelled by radiative feedback, instead remaining trapped in the GMC by the large-scale B-field, producing a nearly one-dimensional flow of gas along the field lines. The trapped gas has a turbulent and magnetic topology that differs from that of the first phase and strongly suppresses gas accretion onto protostellar cores, reducing their masses.
We speculate that this star formation bimodality may be an important ingredient to understand the origin of multiple stellar populations observed in massive globular clusters.
\end{abstract}


\listoffixmes{}

\section{Introduction}

Giant Molecular Clouds (GMCs) are known to be the primary sites of star formation in galaxies owing to their status as the principal reservoirs of the molecular hydrogen supplies of the interstellar medium (ISM) \citep{McKee:1997,Williams:2000}. Despite intensive study and simulation, the mechanisms that govern Star Formation (SF) in these clouds are still not fully understood \citep{Nakamura:2011, Dale:2015, Krumholz:2019, Kim:2021} A complex interplay of turbulent motion, gravitational interaction, stellar feedback, and magnetic fields combine to produce an intricate and hierarchical structure \citep{McKee:2007}. In such a model, small changes to parameters, some of which are poorly constrained by observation, can have large impacts on the stellar populations produced by these clouds \citep{Dale:2015, Krumholz:2019}.  One influence of particular interest is the effect of the magnetic field of the host galaxy on its GMCs. 

Magnetic fields are ubiquitous in galaxies and the ISM and are known to significantly impact gas behaviour \citep{Beck:1996,Rodrigues:2015}. This extends to the scale of GMCs, which are also often observed to be strongly magnetized \citep{Troland:2008,Falgrone:2008,Crutcher:2010}. Generically, it is widely recognized that magnetic fields have a particularly important role at small scales, where the gas reaches high densities. This can be simply understood as, in ideal magneto-hydrodynamics (MHD), magnetic field lines are transported by the fluid elements, resulting in increased magnetic field strengths in higher-density regions. This produces a strong influence on the star formation process, which is controlled by small, dense filaments and protostellar cores \citep{Hennebelle2019}. In principle, the magnetic field can even prevent the collapse of dense cores in molecular clouds (when the cloud becomes sub-critical) or the formation of rotation-supported circumstellar disks via magnetic breaking \citep{Joos2012}. 

It is convenient to parameterize the dynamical impact of the magnetic field in a molecular cloud of mass $M$ in terms of the dimensionless ratio  $\mu \equiv M / M_\Phi$, where $M_{\Phi}$ is the magnetic critical mass defined as the mass at which the pressure from the magnetic energy, ${\cal B}$, balances the gravitational binding energy, ${W}$, of the cloud (see \S~\ref{ssec:mtophi} for further elaboration on our method of computing this ratio).

Observational measurements of the magnetic field strength in molecular clouds come principally from Zeeman effect observations. Such studies observe average values of $\mu ~= 2-3$ and upper extremes of $\mu ~= 5-6$ \citep{Troland:2008,Crutcher:2010}. Further, such surveys also suggest the existence of a large number of molecular clouds with $\mu$ ratios approaching or even below unity \citep{Crutcher:1999,Falgrone:2008}. Geometrically, CO polarization observations have indicated that GMC magnetic fields tend to be fairly uniformly oriented on large scales, with directions typically correlated with the galactic magnetic fields \citep{Li:2011}. The observational presence of collimated magnetic fields with field strengths strong enough to provide considerable contributions to cloud support provides a persuasive impetus to study the SF behaviour of strongly magnetized GMCs. 

Previous simulation work, such as that presented in \cite{Kim:2021}, has attempted to probe this higher field strength regime. They observe significant influences from strong magnetic fields, including significant alignment of filaments perpendicular to the magnetic fields and strongly anisotropic gas motion. Additionally, they find prolonged star formation timescales for strongly magnetized clouds, but suppression of the total SFE. Many simulations of extremely sub-critical clouds, reviewed in \cite{Hennebelle2019}, showed similar suppression of the SFE and anisotropy in gas motion \citep{Nakamura:2011}. The apparent influence of magnetic field strength on star formation prompts further investigations into these influences. 

In this paper, we explore the effects of global magnetic field strengths on star formation at the scales of molecular clouds. We build off of previous work on the subject by \cite{HeRG:2019,HeRG:2020}, which investigated star formation in a large suite of turbulent molecular clouds, spanning a variety of masses and mean densities. In all of these simulations, the magnetic field support was held at a fixed ratio with respect to the turbulent support. The authors discovered that the clouds produced a stellar population consistent with a Kroupa \citep{Kroupa2002} initial mass function (IMF) in the high mass regime, once shifted by a constant multiple to account for the sub-grid-resolution fragmentation of the sink particles. It was further discovered that the IMF was self-similar, with stars formed at any point during the star formation period in the simulation obeying the same relative mass distribution. Lastly, the simulations in this work were found to form stars with an SFR consistent with a Gaussian distribution with respect to time. The star formation timescales, derived from the widths of these Gaussian SF histories, were found to depend principally on the sound crossing time of the clouds.

The simulations performed in \cite{HeRG:2019} include an imposed magnetic field defined by $v_A = 0.2 \sigma_{\rm turb}$ aligned in a uniform orientation. This results in a mass-to-flux ratio $\mu=5.1$ and an average field strength of approximately $10 \mu G$.

This magnetic field strength is physically reasonable, as previously discussed surveys have provided robust evidence of clouds with similar field strength and mass-to-flux ratios. However, clouds with $\mu ~= 5$ tend to fall on the lower B-field edges of such surveys.
Observations suggest that realistic samples of molecular clouds are likely to have stronger magnetic fields than those modeled in this previous work, with average $\mu$ values of $2-3$, ranging down to the region of unity. The uniform orientation of this magnetic field is consistent with previously discussed observational evidence \citep{Li:2011}.

We further note that in the previous simulations by \cite{HeRG:2019}, the magnetic effects are observed to be completely dominated by turbulent effects, leading to little impact on the overall cloud dynamics. The initial magnetic field geometry is dispersed in short order by turbulent gas motion several megayears before star formation begins. As such, the initial magnetic field geometry was found to have no discernible effect on the star formation process in these clouds. 

While the range of cloud masses and mean densities explored in \cite{HeRG:2019} is relevant to a large fraction of high-redshift and local star-forming GMCs, these simulations did not explore the full range of the realistic parameters for the magnetic field strength. Hence, it is possible that the influence of magnetic fields on the development and dynamics of a significant fraction of typical clouds was underestimated. In this work, we expand upon this earlier simulation suite to probe the higher magnetic field regime and determine what influence these more substantial external fields may have on cloud evolution. 

This paper is organized as follows. In \S~\ref{sec:sim} we introduce our methods and simulations, and in \S~\ref{sec:res} we present the results of our simulations. We interpret these results and discuss their implications in \S~\ref{sec:disc} and conclude in \S~\ref{sec:sum}. 

\section{Simulations and Methods}\label{sec:sim}

We conduct radiation-MHD simulations of molecular cloud collapse that resolve individual massive stars. Our simulations are performed using the grid-based adaptive mesh refinement (AMR) MHD code \ramses{} \citep{Teyssier2002,Bleuler2014}. Radiation transfer is implemented using a moment method with M1 closure \citep{Rosdahl2013}. The ionizing photons emitted from stars interact with neutral gas; we keep track of the ionization chemistry of hydrogen and helium, but we do not include the chemical evolution of the molecular phase. Heating from photoionization and cooling from hydrogen, helium, metals, and dust grains are implemented (see \citealt{Geen2017} for details). Cooling below 10~K is shut down to keep the temperature floor at 10K. 

Mesh refinement is applied to the whole domain adaptively to ensure that the local Jeans length, $L_J = c_s \sqrt{\pi/(G \rho)}$, is resolved by at least 10 grid points at any time and any location. When the number density reaches $n_{\rm sink} = {3.0} \times 10^6\  \pcc$, defined such that the local Jeans length equals 5$\times$ the grid size at the maximum refinement level (14), a sink particle is placed to represent a single star or a small cluster of stars. It is shown in \cite{HeRG:2019} that these sink particles represent prestellar cores which have a mass function that matches the empirical stellar IMF when shifted to the lower-mass end by $\sim 40\%$. This recipe is supported by zoom-in simulations that resolve the collapse of individual prestellar cores \citep{He2022x}.

Ionizing photons are emitted from stars and heat the gas. The ionizing luminosity is calculated through a fit to the data from \cite{Vacca1996}.
We use the following formula as the ionizing luminosity of a sink particle with mass 
$M_s$: $S = 9.6 \times 10^{48} (0.4 M_s/27~M_\odot)^{1.87} \ {\rm s}^{-1}$. 
This is an extension of the high-mass ($\gtrsim 30 \msun$) end of a fit to the data from \cite{Vacca1996} into the lower-mass end.
The excess of ionizing photons from stars below 10 - 30 \msun{} is used to compensate for the lack of protostellar feedback. This recipe is proved effective in reproducing the star formation efficiency and stellar initial mass function from observations \citep{HeRG:2019}.  
Further simulations with more realistic feedback mechanisms are left for future work. 

The motion of the sink particles is determined by combining direct N-body integration between the sinks and between the sinks and the gas based on the particle mesh method. A softening length of $2 \Delta x_{\rm min}$, where $\Delta x_{\rm min} = 1000$ AU is the spatial resolution, is set to avoid singularities.

\subsection{Mass-to-flux ratio calculation}\label{ssec:mtophi}
 
An important metric for evaluating the dynamical influence of the magnetic field strength in a cloud is the mass-to-flux ratio $M/\Phi_B$. An equivalent definition is used as
its ratio to the critical value, $\mu \equiv (M/\Phi_B) / (M_\Phi/\Phi_B)$, where $M_{\Phi}$ is the magnetic critical mass, 
the mass at which the pressure from the magnetic energy, ${\cal B}$, balances the gravitational binding energy, ${W}$, of the cloud.
We derive this relation from the study of a spherical cloud with uniform density threaded by a uniform magnetic field. The total gravitational binding energy of such a cloud is ${W}=-3/5 GM^2/R$ and the magnetic energy is ${\cal B}=B^2R^3 / 6 = \Phi_B^2 / 6 \pi^2 R$, where the magnetic flux $\Phi_B \equiv \int B_{\perp} dS = \pi R^2B$. We can calculate the magnetic critical mass
\begin{equation}\label{eq:mphi}
M_{\Phi} = c_{\Phi} \frac{\Phi_B}{G^{1/2}},
\end{equation}
where $c_\Phi = 0.17$ for the aforementioned geometry.

However, this formulation of the magnetic critical number $\mu$ becomes less reliable for systems that depart from uniform density and physical symmetry. To provide a more robust formulation that remains accurate for a general inhomogeneous and/or asymmetric mass distribution, we utilize the energies directly, noting that
\begin{equation}\label{eq:eWeB}
  \frac{|{W}|}{{\cal B}} = \frac{18 \pi^2}{5} \frac{GM^2}{\Phi_B^2} = \frac{M^2}{M_{\Phi}^2} = \mu^2.
\end{equation}
The second equal sign holds for the uniform spherical geometry. 
We thus adopt a more precise definition of the mass-to-flux ratio, $\tilde{\mu} \equiv \sqrt{|{W}|/{\cal B}}$, to account for the inhomogeneity of the density and magnetic field distribution. This method also has the benefit of being easy to calculate numerically at any point during the cloud evolution. For an ideal uniform sphere, $\tilde{\mu} = \mu$. 
For a more centrally concentrated geometry (e.g., non-singular isothermal sphere), the equivalent geometrical factor $c_{\Phi}$ is up to 70\% higher and $\tilde{\mu}$ is 40\% lower \citep{Heinprep}.

\subsection{Simulations}\label{ssec:sims} 
Using these methods, we perform a set of three simulations. These runs are extensions of the M-C cloud from \cite{HeRG:2019}, and thus all feature identical gas mass, gas density profile, metallicity, and initial turbulent velocity field. Specifically, this corresponds to a mass of $43,300 M_{\odot}$ and a radius of $10$ pc. 
The clouds have an initial density profile of a non-singular isothermal sphere in hydrodynamic equilibrium with a turbulent velocity field following a Kolmogorov power spectrum ($P(k) \sim k^{-5/3}$) with random phases. We let the cloud relax for a period of time to allow turbulence to fully develop by halving self-gravity before turning on full gravity. At the time when full gravity is turned on, all three clouds have consistent average densities of $\sim 125-150 \ \pcc{}$, typical of Milky Way star-forming molecular clouds \citep{Williams:2000}. See Table \ref{tab:allparam} for a complete set of initial conditions.

Note that the calculation of average density here differs from that in \cite{HeRG:2019} and is a more accurate definition of the true cloud mean density. In \cite{HeRG:2019} the density of the cloud is defined as the average density of the initial isothermal sphere within half of the cloud radius (because the cloud is embedded in a constant density envelope with a radius twice the isothermal sphere radius). This definition returns values 14 times higher than those quoted here, which average over the density of the entire cloud after the cloud has relaxed its turbulent field and the isothermal core is mixed with the lower-density envelope.

The only differences between the 3 simulations are the strength of the global magnetic field. In Run 0, which we refer to as the ``low field'' or ``fiducial'' run, the magnetic field strength is the same as in previous work, with $\mu = 5.1$. Run 1 increases magnetic field strength by a factor of 2, producing $\mu = 2.5$; as this is within the range of typical mass-to-flux ratios found in observation, it is referred to as the ``intermediate'' or ``average'' run. Finally, Run 2's magnetic field strength was increased by a factor of 5 over the fiducial, resulting in $\mu = 1.0$. This run is referred to as the ``high field'' run. This $\mu$ value represents a critical cloud, where the magnetic support nominally balances gravitational forces. In practice, clouds with this support are still able to collapse, a somewhat counter-intuitive effect resulting from differences between magnetic support and more typical thermal or turbulent support. For uniformly oriented magnetic fields, gas may freely flow along the magnetic field lines, and thus a cloud is only supported against gravitational collapse along the axes perpendicular to the magnetic field direction. Such critical, and even sub-critical, clouds have been observed in nature \citep{Crutcher:1999,Falgrone:2008}. A list of simulation parameters for these three runs is included in Table~\ref{tab:initparam}.
\begin{table}
    \centering
    \begin{tabular}{ccc}
    Name & $B_0$ ($\mu$G) & $\mu_0$ \\
    \hline 
     Fiducial B-field Run & 11.7 &  5.2 \\
     Intermediate B-Field Run & 23.4 & 2.6 \\
     High B-Field Run & 58.4 & 1.0\\
    \end{tabular}
    \caption{A table of the initial magnetic field strengths and $\mu$ values for the 3 simulations presented in this work.}
    \label{tab:initparam}
\end{table}
Note that the magnetic field strengths presented in Table \ref{tab:initparam} are averages over the entire cloud. The initial conditions of our simulation preserve a $B \propto \rho^{1/2}$ scaling relation across a cross-section of the cloud passing through the centre and perpendicular to the field direction. As a result, the magnetic field strength is strongest in the centre of the isothermal core, dropping in magnitude towards the edges of the cloud. The magnetic field strength in the ambient medium is initialized in equilibrium with the field strength at the edge of the cloud and is typically $\sim 15\%$ lower than the average strength within the cloud, reported in Table \ref{tab:initparam}.

In addition to representing realistic $\mu$ values for GMCs, the magnetic fields in these simulations sample a range of realistic large-scale galactic field strengths in likely star-forming regions. The fiducial B-field simulation roughly corresponds to the ambient field strength of $6-10~\mu G$ in the local solar vicinity \citep{Crocker:2010}. The intermediate B-Field run is consistent with the $20-30~\mu G$ field strength in spiral arms and bars \citep{Beck:2015}. The high B-Field simulation is within the range of $50-100~\mu G$ observed in some starburst galaxies and within the central regions of spirals, including the Milky Way \citep{Crocker:2010,Adebahr:2013,Beck:2015}.

\section{Results}\label{sec:res}

\begin{figure*}
\includegraphics[width=\columnwidth]{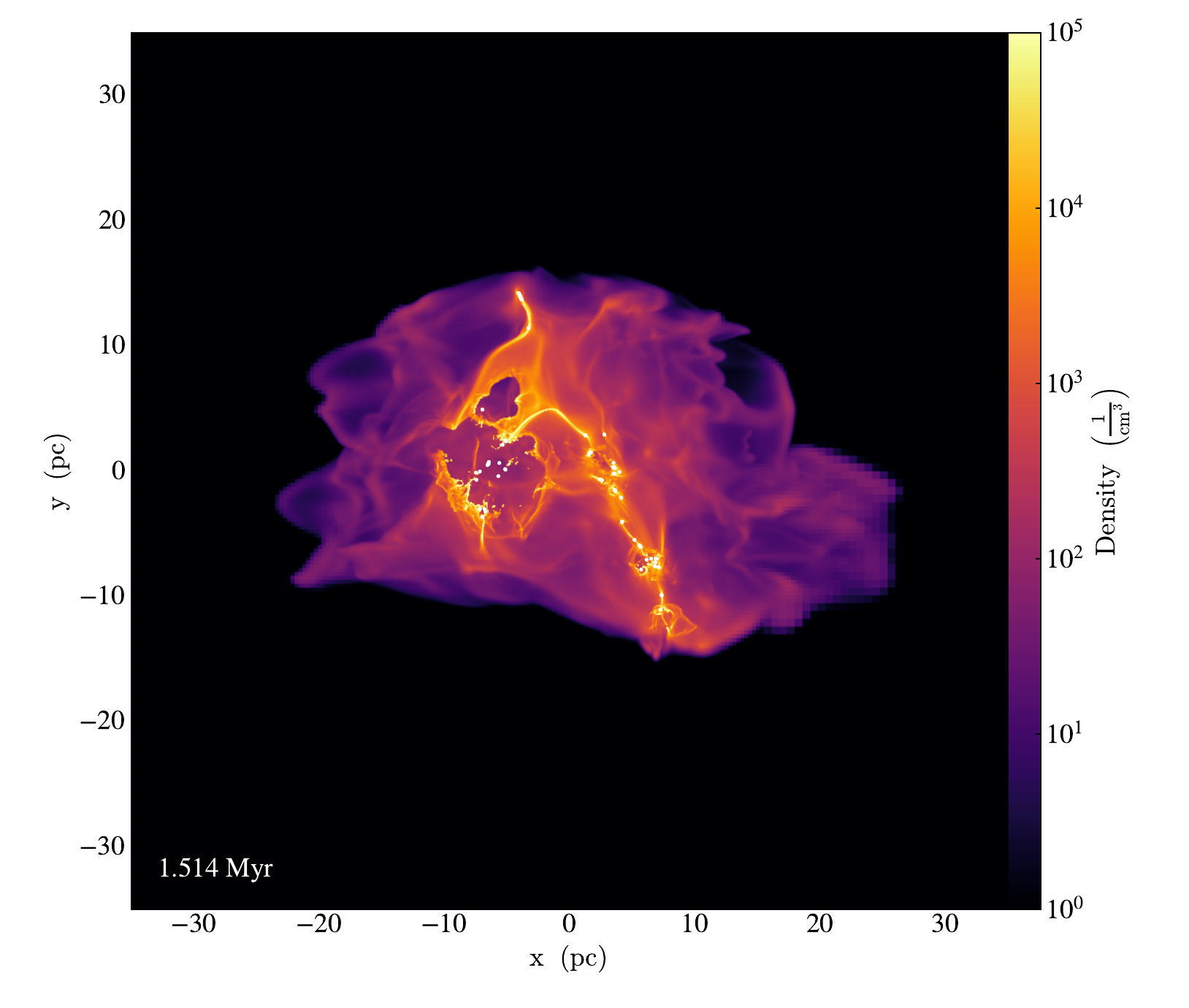}
\includegraphics[width=\columnwidth]{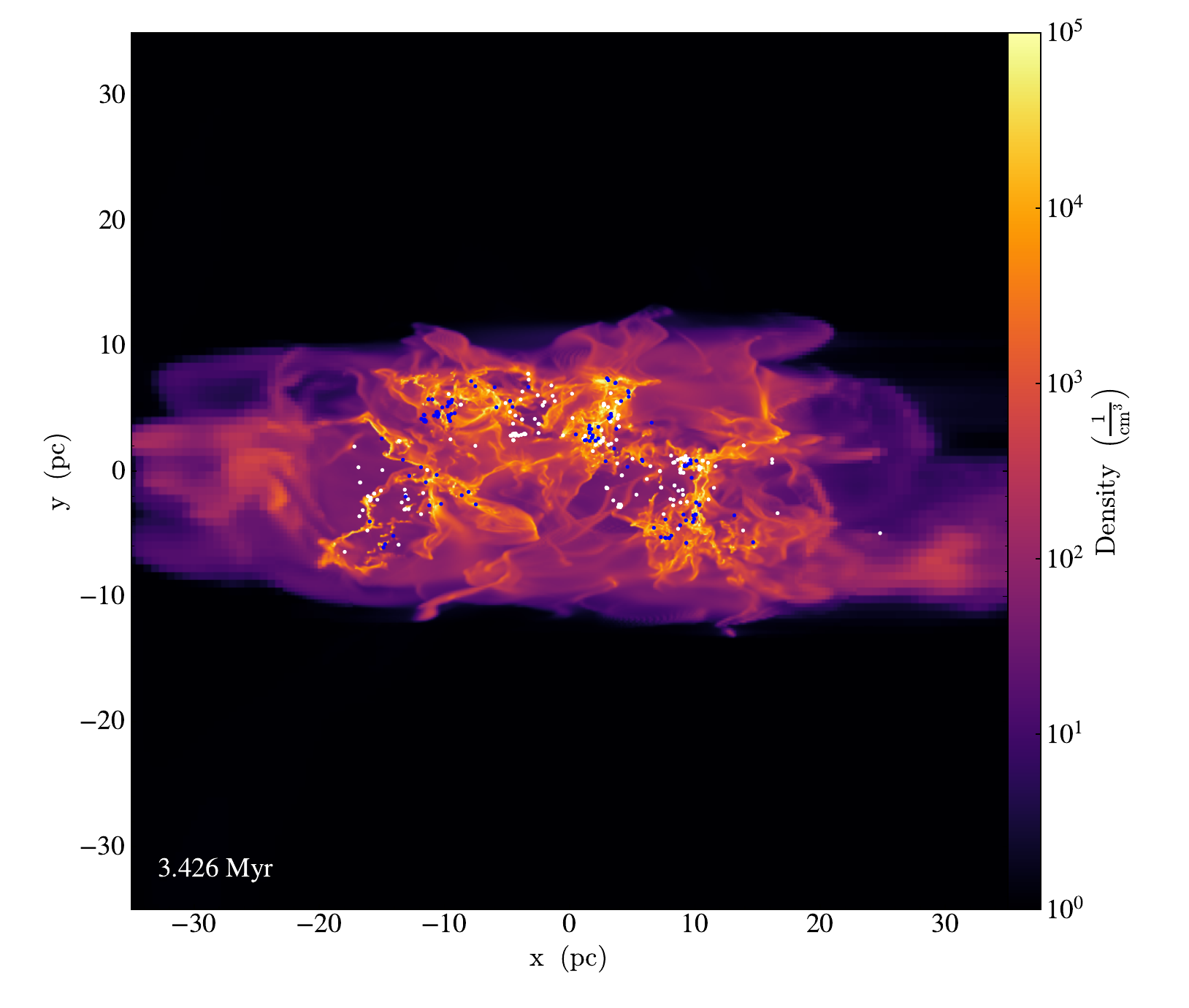}
\caption{Density projections of the fiducial (left) and high magnetic field (right) simulations. The initial magnetic field is oriented parallel to the x-axis in both clouds. Stars are marked with dots; the distinction between the white and blue colours of some stars is discussed in \S~\ref{ssec:sfe}. Note the visible confinement along the y-axis and elongation along the x-axis in the high B-field run. }
\label{fig:confine}
\end{figure*}

The strength of the initial magnetic field has a clear effect on the large-scale geometry of the cloud, as shown in Fig. \ref{fig:confine}, comparing snapshots of the fiducial run to the strong magnetic field case. When compared to the fiducial run, the intermediate and high B-field clouds were significantly confined in the directions perpendicular to the imposed magnetic field. In the case of a strong magnetic field run, the cloud's extent in the dimension parallel to the applied field was initially consistent with the behaviour of the fiducial run, but the extent of the cloud in the 2 perpendicular dimensions was confined within $\pm$10pc throughout the entire star formation period. This constrained behaviour was also observed in the intermediate B-field run, but to a weaker degree. This confinement along 2 axes is consistent with the anisotropy of magnetic support discussed previously.

Interestingly, this confinement persisted in the high B-field run after star formation began, despite the presence of strong photoionizing UV bubbles from stellar feedback. In the fiducial run, these hot regions of gas rapidly expand outwards, isotropically ejecting gas from the cloud and quenching star formation. Several examples of these bubbles, early in their expansion, can be seen in Fig.~\ref{fig:confine} in the fiducial cloud projection, particularly at (X=-10 pc,Y= 0 pc). In the presence of a strong magnetic field, gas ejection was almost completely suppressed in the direction perpendicular to the magnetic field. As a result, the broadly spherical isotropic expulsion of gas seen in the fiducial run was replaced by a one-dimensional ``pipe-like'' gas ejection mechanism. This restriction resulted in a slower overall ejection of gas from the cloud.

\begin{figure*}
\includegraphics[width=\columnwidth]{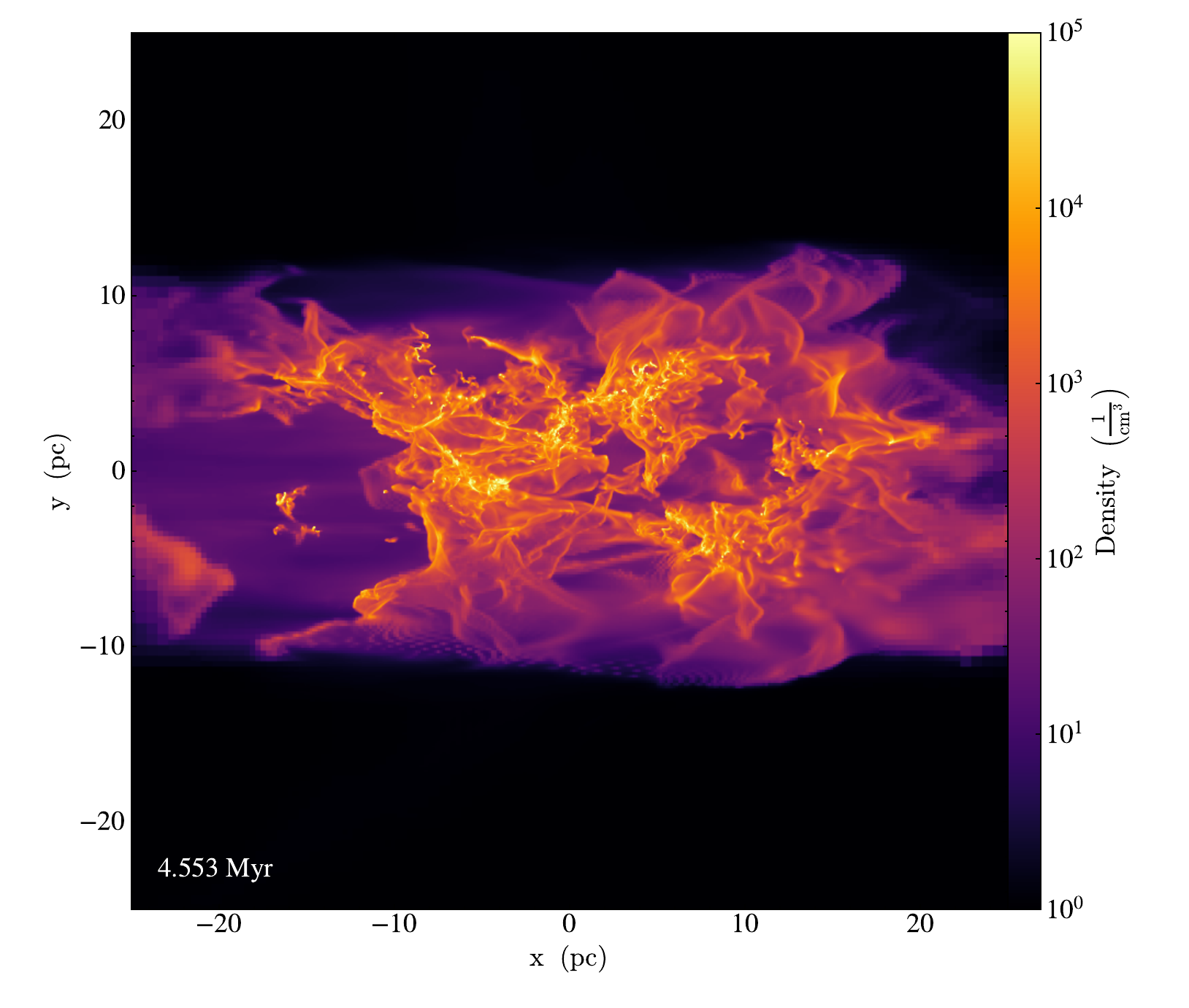}
\includegraphics[width=\columnwidth]{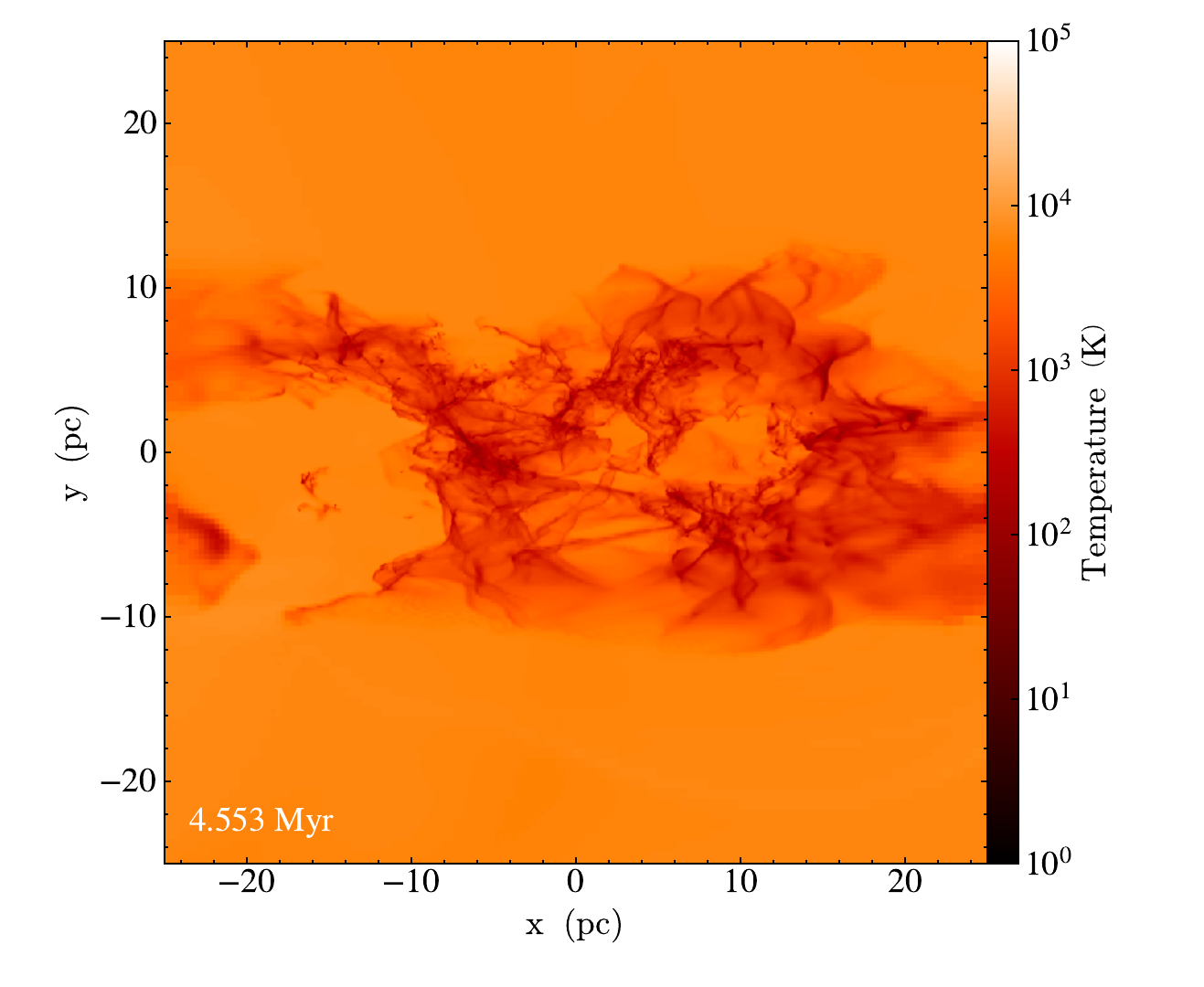}
\caption{
Narrower view of the central region of the high B-field cloud at $t=4.3$~Myrs, roughly at the beginning of the second phase of star formation. A density projection is shown on the left, and a mass-weighted temperature projection is shown on the right. Note that the extent of the cloud in the y direction is remarkably unchanged with respect to earlier snapshots. Further, this region has a distinct character of turbulence with respect to the weakly magnetized cloud, consisting of many thin, chaotic filaments of cold, dense gas.}\label{fig:frames1}
\end{figure*}

As a result of this confinement, significant quantities of cold, dense gas remained in the high-field cloud for a much longer period than in the fiducial simulation. This is especially noticeable in the central region of the cloud, shown in Fig.~\ref{fig:frames1}, where nontrivial quantities of gas linger for more than 10 Mrys after the formation of the first stars. This stands in contrast to the fiducial run, where all gas is ejected within approximately 3 Myrs. In the intermediate run, gas expulsion takes around 6 Myrs. Of further interest is the behaviour and structure of the gas in this central region, visible in the density projection in the left panel of Fig.~\ref{fig:frames1}. Initially, this region was characterized by several large filaments, with very similar forms to those formed in the fiducial cloud. However, following the collapse of these large-scale filaments, the gas in the central region shifted to display a highly chaotic structure, characterized by a web of small, interconnected filaments. The right panel of this figure, a density-weighted temperature projection, shows that many of these turbulent filaments are composed of cold gas, suitable for star formation. 

Probable physical sources and influences on these geometric effects are discussed in Section \ref{sec:disc}. However, the presence of differences in the cloud geometries of the fiducial and more strongly magnetized clouds prompts an investigation into any potential differences in their star-forming histories.

\subsection{Dependence of Star Formation Efficiency on Magnetic field strength}\label{ssec:sfe}

\begin{figure*}
\includegraphics[width=\columnwidth]{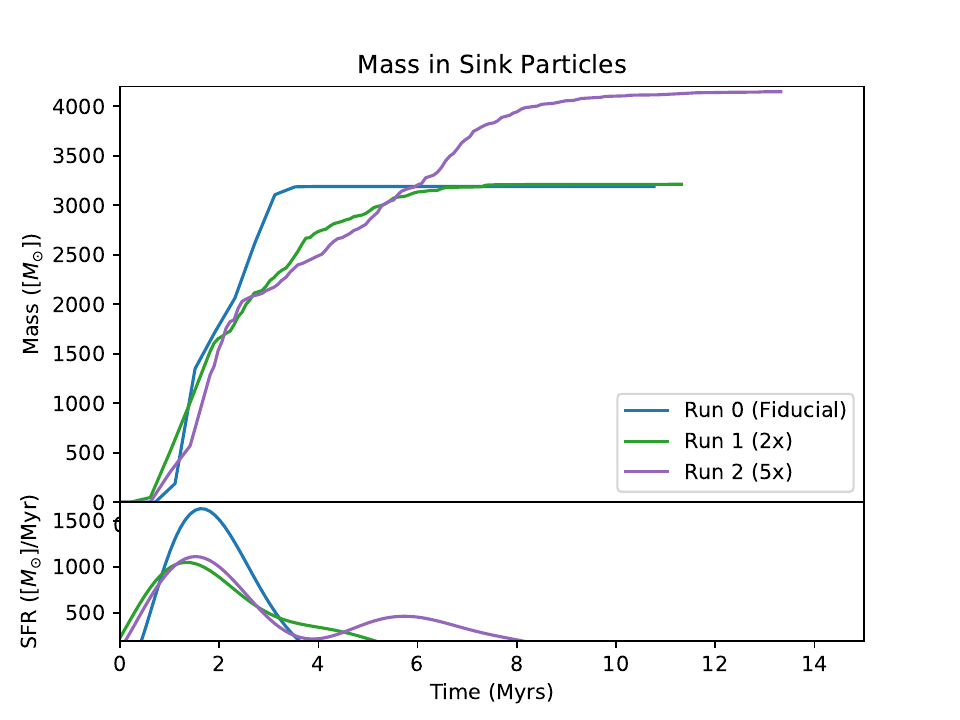}
\includegraphics[width=\columnwidth]{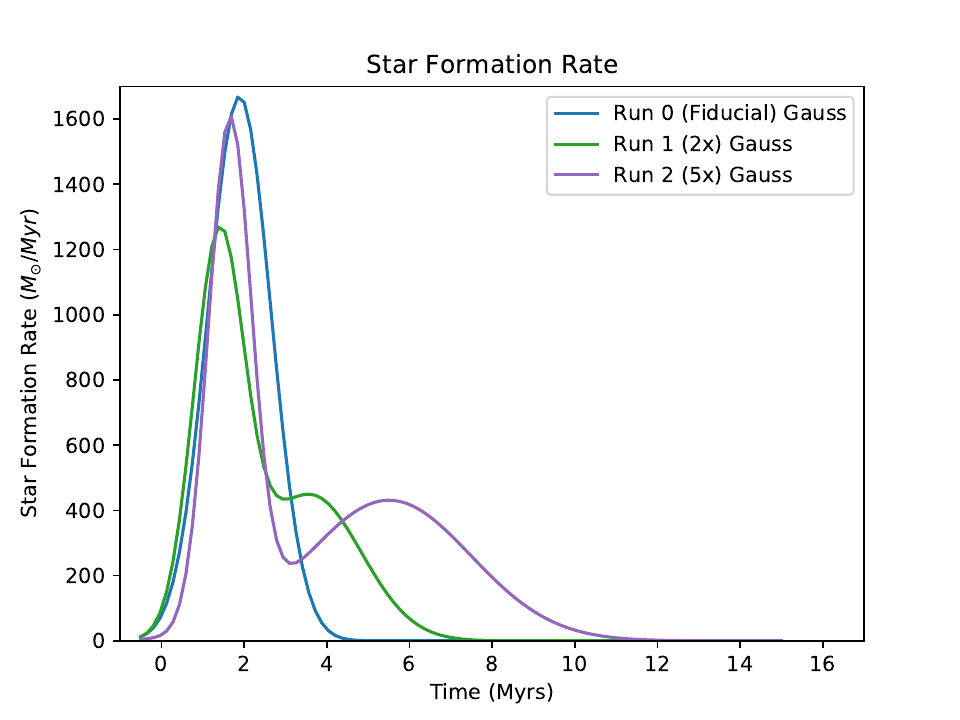}
\caption{Left: The cumulative mass in sink particles, $M_*(t)$, as a function of time for the three simulations in this study (see legend). The insert at the bottom shows the SFR as a function of time for the same simulations, obtained by taking the time derivative of a 5-degree polynomial fit to $M_*(t)$ (in order to smooth out the noise of the SFR). Right:  The  best fit using Gaussian functions to the star formation rate as a function of time for the three simulations in this study (see legend). We use one Gaussian (for the fiducial run) or the sum of two Gaussians (for twice or 5 times the B-field strength) for the functional form of the SFR and fit the integral to $M_*(t)$.}
\label{fig:sinkmass}
\end{figure*}

The most interesting effects of the magnetic field are upon the resultant stellar population. The left panel of Fig.~\ref{fig:sinkmass} shows the total mass in sink particles plotted versus time; the slope of this curve being the instantaneous star formation rate (SFR). In all 3 runs, star formation starts promptly after the cloud relaxation period and proceeds in a relatively Gaussian manner, consistent with \cite{HeRG:2019} and \cite{Bate2019}, although the latter only provides data for the evolution of the SFR for the initial $\sim 1.5~t_{ff}$.
In the baseline cloud, total stellar mass eventually asymptotes, as star formation tapers off after approximately 3 Myrs. In the intermediate field case, this cessation of star formation begins earlier, but is significantly prolonged, resulting in a longer star formation period, but a nearly identical total star formation efficiency (SFE). 

In the high field run, the star formation process undergoes a notable change at 2 -- 3 Myrs. After initially tracking the baseline run, the SFR decreases markedly. This second, reduced SFR is relatively constant for another several Myrs, before also quenching. As a result of this significantly longer total star formation period, the high magnetic field run displays a higher total star formation efficiency. Such behaviour is unusual, as previous simulations at fiducial B-field strengths find star formation histories consistent with the cumulative distribution function (CDF) of a Gaussian in time. Large departures from this functional form, as is seen in the strongly magnetized simulations presented here, have not been observed in these previous simulations. Further, the simulated cloud is large enough that this result would have to come from a global shift in cloud development, as any localized feature or effect would lack the statistical power to skew the overall SFR so markedly or for such a long period. The overall star formation results and SFE values are included in Table \ref{tab:starvals}.

To further investigate this SFR shift, as well as the prolonged star formation quenching timescale, a more detailed quantification of the SFR is required. A point-by-point finite difference method is highly vulnerable to random sampling noise, necessitating a smoother form. To obtain such a differentiable, continuous form for the star formation, a high-order polynomial in time was fit to the total sink mass data. This fit captures the overall forms and trends of the data with high fidelity and provides a qualitative overview of the star formation behaviour. The time derivative of this fit is the SFR and is shown at the bottom of Fig.~\ref{fig:sinkmass} (left). These fits are not strictly physical, but provide a reference for the general evolution of the SFR. More physically meaningful fits of the star formation history are provided later in this work. 

Consistent with previous work, the fits reveal the fiducial run's SFR to be fairly consistent with a single Gaussian, and the total mass in sink particles tracks the CDF of a Gaussian. The other runs instead display well-defined but more complex forms. The intermediate run appears to also be initially Gaussian, but includes a shelf at late times, consistent with the previously observed extended time for star formation quenching, as compared with the fiducial run. Even more interesting is the high B-field run, which displays two distinct peaks: an initial peak temporally consistent with the star formation period of the baseline run, and a second peak several megayears later. The prominence of these features in the higher field strength runs prompts a more thorough and physically rigorous analysis of the star formation rate, to better understand the strength and sources of these unusual secondary features.

To accurately model the multi-feature SFR progression, the total stellar mass was fit using the CDF of the sum of two Gaussians, yielding an SFR modelled by the sum of two Gaussians:
\begin{equation}
    f(t)=\sum_{i=1}^2 \frac{A_i}{\Delta t_i \sqrt{2\pi}}\exp{\left(-\frac{(t-t_i)^2}{2\Delta t_i^2}\right)}.
    \label{eq:gauss}
\end{equation}
This method produced a robust fit to the data and consistency with the qualitative form of unbiased polynomial fits. A plot of the SFRs produced by this method is shown in the right panel of Fig.~\ref{fig:sinkmass}. Best fit parameters for these Gaussians are provided in Table \ref{tab:gaussfits}. We note that our limited sample size makes reliable quantification of the goodness of fitting difficult, and thus errors on the fitting parameters are likely much higher than those yielded by the covariance of the fitting parameters quoted in the table. Nevertheless, the correspondence of these fits with the unbiased polynomial fits and the lack of any noticeable systematic deviations from the data indicate that they serve as a reasonable model of the SFR behaviour. It is visually evident that the higher magnetic field runs are best modelled with a second period of star formation.

The uncertainties in the fitting process and the stochasticity inherent in the simulations preclude any precise claim about the relative peak SFR between the different runs. We can, however, note that all three simulations display initial peaks at the same approximate times, with comparable peak SFRs.

The primary differences between the runs lie in the secondary peaks. In the baseline run, this peak is entirely absent, but it is a major contributor to the overall stellar population in the higher B-field runs. Overall, this second epoch has a much lower SFR than the initial epoch but continues for a much longer period. The fit amplitude coefficients ($A_1$  and $A_2$) are normalized to equal the total mass of sink particles formed in each phase. Inspection reveals that approximately half of the total stellar mass produced by the non-fiducial clouds results from this second epoch of star formation. In both high B-field clouds the peak SFRs in the second epoch are comparable, although this peak is more delayed in the highest field run. 

\begin{figure*}
\includegraphics[width=\columnwidth]{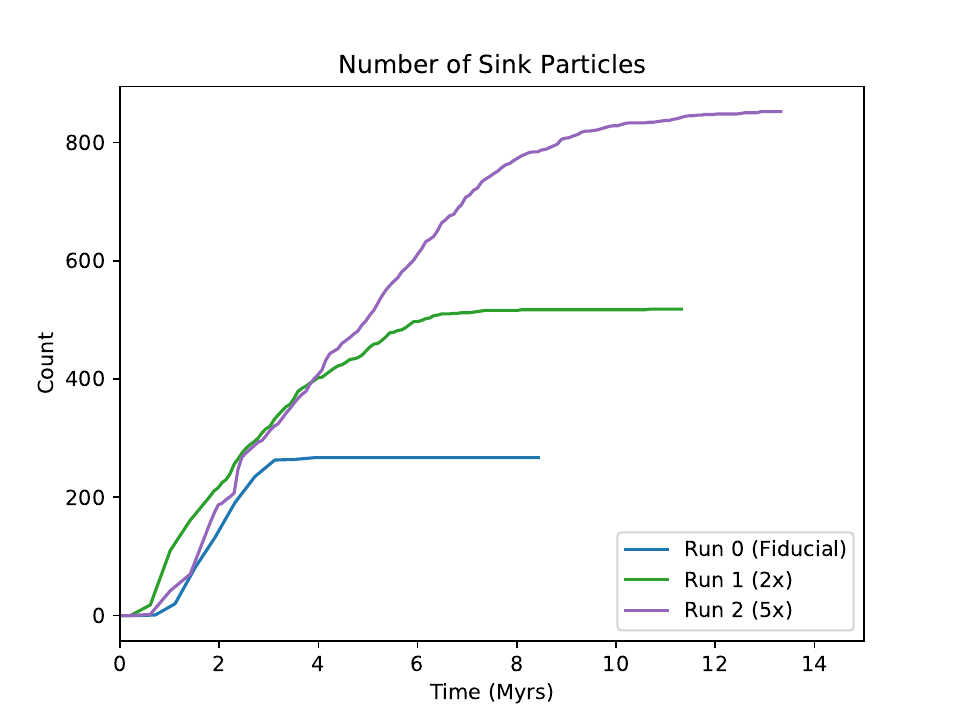}
\includegraphics[width=\columnwidth]{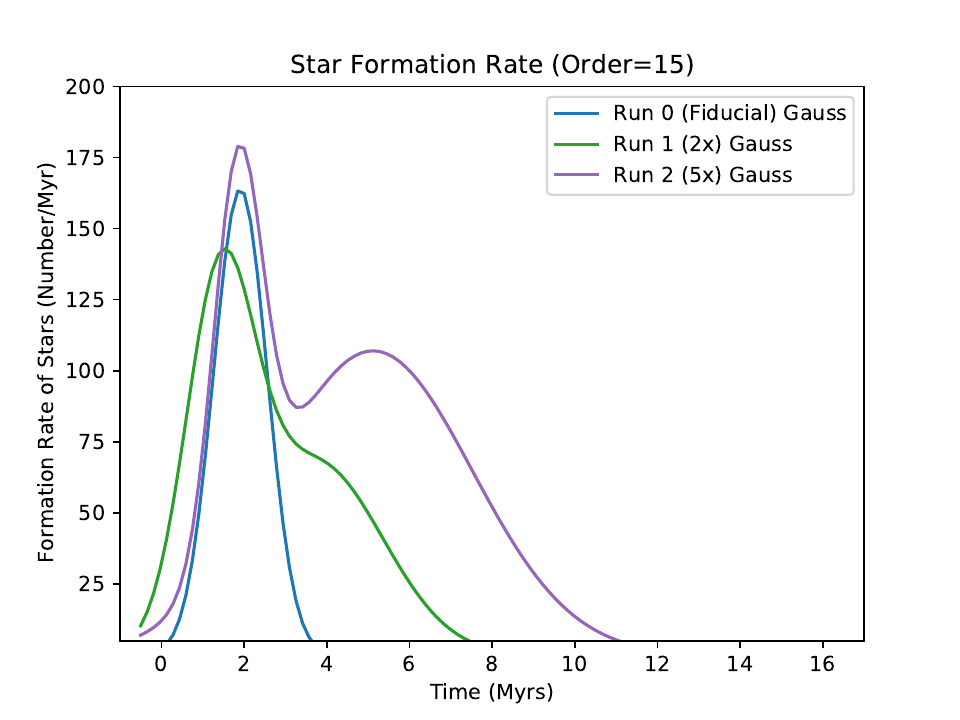}
\caption{Left: The number of sink particles as a function of time for the three simulations in this study (see legend). Right: Sink formation rate as a function of time for the same three simulations. The lines are equivalent to those in \protect{Fig.~\ref{fig:sinkmass}} but for the number of sink particles instead of the total mass in stars.} 
\label{fig:sinknum}
\end{figure*}

We further apply the same methodology to the number of sink particles in the three runs. Fig.~\ref{fig:sinknum} shows the number of sink particles versus time for each of the three simulations, along with the numerical formation rate of sink particles. Note in the right panel that the same bi-modality present in the SFR is observable in the numerical star formation rate (NSFR) -- the rate at which stars form as measured by number, rather than mass -- lending additional support to the two-epoch model described above. There are, however, several notable differences. 

As can be seen in the left panel of Fig.~\ref{fig:sinknum}, the final number of sink particles in each of the runs increases with an increasing magnetic field. The intermediate magnetic field produced approximately twice as many stars as the fiducial run, while the highest field run produced around three and a half times as many. This is notable as the total mass in sink particles for the fiducial and intermediate runs have the same final mass in stars, while the high magnetic field run has only ~30\% more mass in stars. This implies a significant difference in the average stellar mass between the three runs. Further, the secondary peak in the NSFR plot (right panel of Fig.~\ref{fig:sinknum}) has a location and duration consistent with the preceding SFR plot; however, the relative peak heights differ. The ratio of the primary and secondary peak NSFRs is significantly closer to unity than the ratio of SFRs, implying that the stars formed in the second star-forming epoch have a considerably lower average mass than those formed in the first epoch. This observation prompts a detailed analysis of the characteristics of the two stellar populations present in both the enhanced magnetic field runs. This topic is explored in Section \ref{ssec:imf}; a table of summary values for the stellar populations of each run is provided in Table~\ref{tab:starvals}.

\begin{table*}
	\caption{Table of fitting parameters for the SF relations, defined in accordance with \ref{eq:gauss}. We emphasize that the true uncertainties on these variables are likely significantly larger than fit results would suggest, owing to stochasticity from a limited number of simulations.}
	\label{tab:gaussfits}
	\begin{tabular}{lccccc|ccccccc} %
		\hline
		Sim ID & $A_1$ & $A_2$ &  $t_1$ & $t_2$ & $\Delta t_1$ & $\Delta t_2$ \\
		& $M_{\odot}$ & $M_{\odot}$ & Myr & Myr & Myr & Myr\\
	    \hline
	    Fiducial & $3189\pm7$ & 0 & $1.89\pm.02$ & 0 & $0.762\pm.03$ & 0 &\\
		2x B-field & $1805\pm122$ & $1389\pm123$ & $1.38\pm.03$ & $3.6\pm.15$ & $0.610\pm.04$ & $1.24\pm.10$ &\\
		5x B-field &  $1998\pm30$ & $2141\pm32$ & $1.66\pm.01$ & $5.51\pm.04$ & $0.516\pm.02$ & $1.98\pm04$ &\\
		\hline
	\end{tabular}
\end{table*}

\begin{table}
	\caption{Final stellar population values. Note that the Phase 1 and Phase 2 average masses are derived from fits, owing to the necessity of separating the two phases, while all other values are from unprocessed data.}
	\label{tab:starvals}
	\begin{tabular}{lcccc|cccc} %
		\hline
        Parameter & Fiducial & Intermediate & High B-Field \\
        \hline
        $M_{tot}$ ($M_{\odot}$) & 3189 & 3211 & 4146\\
        $N_{stars} (\#) $ & 267 & 518 & 852 \\
        $M_{avg}$ ($M_{\odot}$) - Phase 1 & 11.94 & 7.43 & 9.89\\
        $M_{avg}$ ($M_{\odot}$) - Phase 2 & -- & 5.06 & 3.32\\
        $M_{avg}$ ($M_{\odot}$) - Overall & 11.94 & 6.20 & 4.87\\
        SFE (\%) & 7.36 & 7.42 & 9.58\\
		\hline
	\end{tabular}
\end{table}

The SF history of these clouds seems to indicate a second period of star formation in the more strongly magnetized clouds, with different SFRs and SF durations, as well as differences in average mass. This is a notable departure from previous simulations, which display single, self-similar star formation periods. Stars with formation times consistent with this second formation period are denoted in Figure~\ref{fig:confine} with blue, rather than white, points.

\subsection{Dependence of the Initial Mass Function on Magnetic Field Strength}\label{ssec:imf}
    
The presence of two formation periods and the apparent discrepancies in the average masses prompt a detailed analysis of the stellar populations formed under each magnetic field environment, best enabled by analysis of the IMF.

We note that the stellar mass functions are produced by scaling the sink particle mass by a factor of 0.4 to account for the core-to-star conversion, i.e. fragmentation of a sink particle into multiple stars. This recipe is motivated by the matching between sink particle mass function and empirical stellar IMF in our previous work \citep{HeRG:2019}. We emphasize that the simulations presented in this work do not possess sufficient resolution to directly simulate prestellar core fragmentation below our resolution limit of 1000 AU. If pre-stellar cores of different masses in our simulations undergo substantially different fragmentation, our derived IMFs could be skewed. The implications of this core-to-star scaling and the study of sink particle fragmentation at high resolution are discussed in ~\S \ref{ssec:prestellarinf}.

\begin{figure*}
\includegraphics[width=\columnwidth]{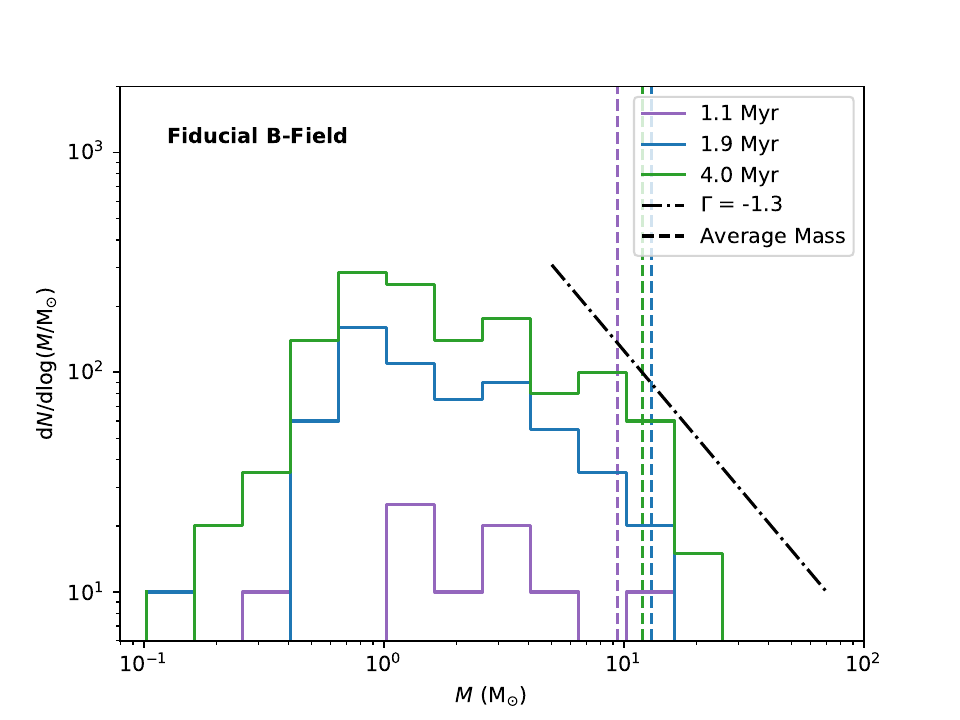}
\includegraphics[width=\columnwidth]{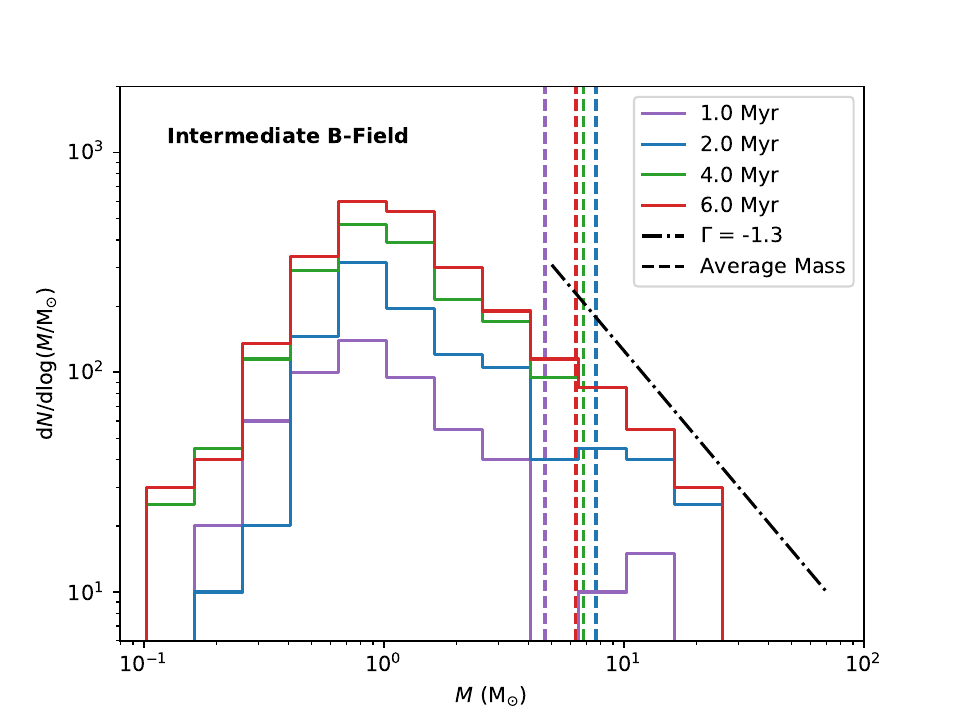}
\caption{
Plots of the mass functions in the fiducial (left panel) and intermediate B-field runs (right panel). The dashed vertical lines denote average mass. Note how in both runs star formation occurs on statistical grounds, and the mass function increases self-similarly. Note, however, the lower average masses yielded in the intermediate B-field run. The dot-dashed line shows, for reference, the Salpeter power-law slope at the high-mass end of the IMF.}
\label{fig:imf1}
\end{figure*}

Fig.~\ref{fig:imf1} shows the stellar mass functions for the baseline and intermediate magnetic field run at a variety of simulation times. 
In the fiducial run, the stellar populations in each mass bin increase fairly proportionally with time, producing a self-similar scaling of the IMF. This, along with the reasonably constant average mass after the first Myr, suggests a fairly statistical star formation description. Further, the higher mass end of the distribution seems to roughly reproduce a Kroupa scaling with a log-scale power-law exponent $\Gamma = -1.3$ \citep{Kroupa2002}. 
That is, stars formed at any point during the cloud evolution obey approximately the same mass distribution, consistent with the results of \cite{HeRG:2019}.

In the intermediate B-field run, no strong departure from self-similarity is observed, although the intermediate run's average mass is notably lower than the baseline and drifts consistently lower after 2 Myrs. This indicates a stellar population with a larger proportion of low-mass stars, and with the proportion of smaller stars increasing over time, though this behaviour is not visibly obvious. It is, however, worth noting that the intermediate B-field run's second phase of star formation overlaps significantly in time with its first phase. Disentangling the stars of these two phases is thus difficult before 4 Myrs, by which time the majority of the stellar population has been formed.

\begin{figure*}
\includegraphics[width=\columnwidth]{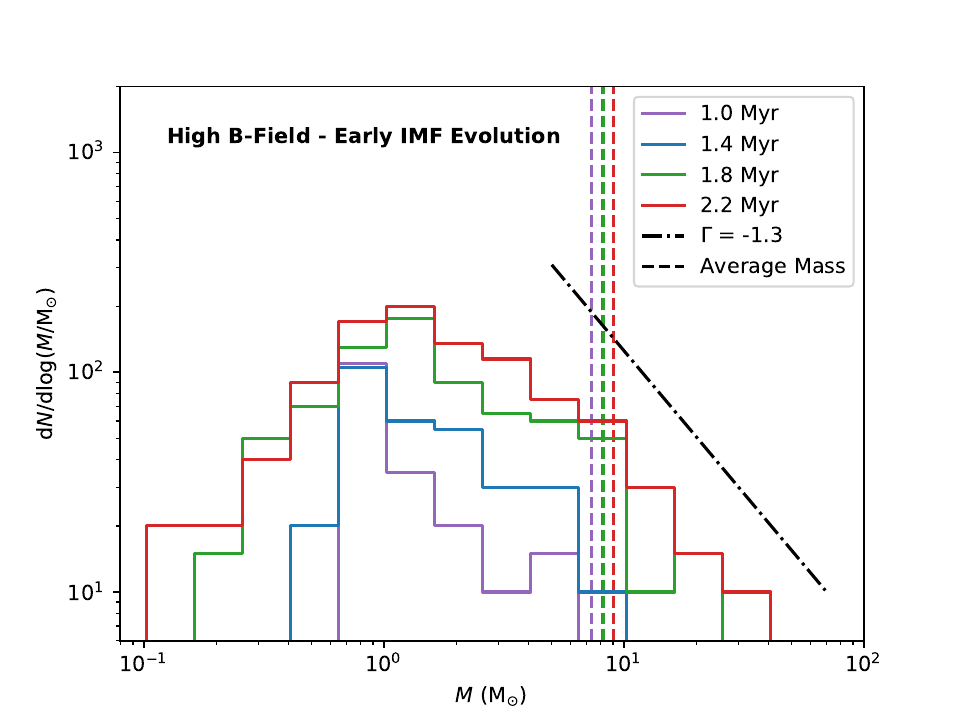}
\includegraphics[width=\columnwidth]{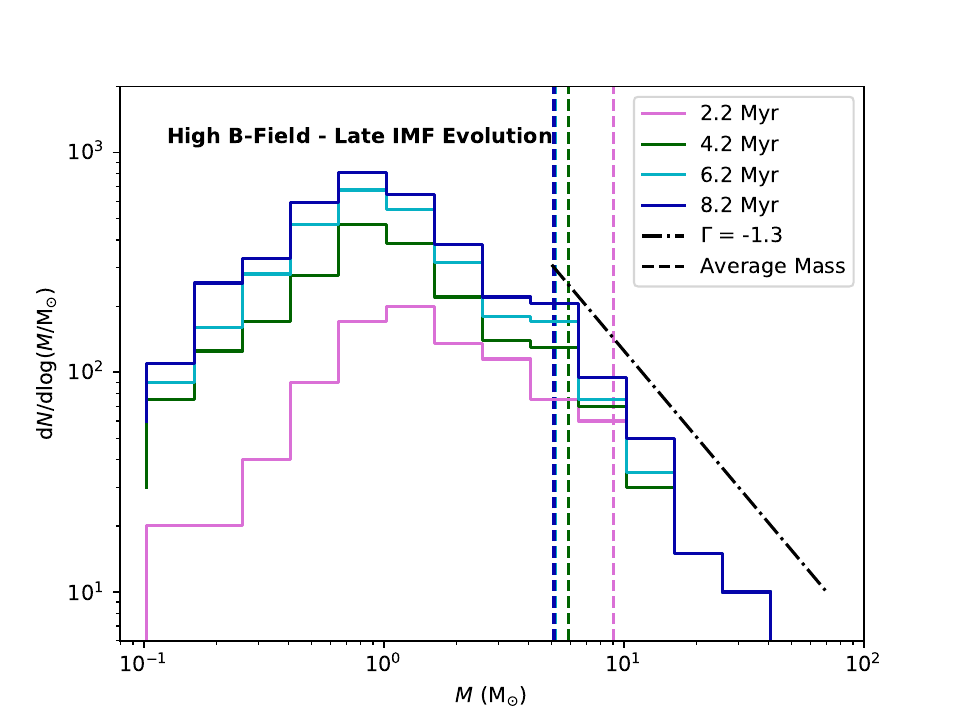}
\caption{Plots of the mass functions for the high magnetic field run before (left panel) and after 2.2~Myrs (right panel). Note that the first phase of star formation follows the same self-similar form as seen in the fiducial and intermediate runs; however, the stars forming during the second phase of SF are predominantly of significantly lower mass, with very few high-mass stars. The dashed and dot-dashed lines have the same meaning as in Fig.~\ref{fig:imf1}.}
\label{fig:imf2}
\end{figure*}

Fig.~\ref{fig:imf2} shows the IMF evolution of the high magnetic field simulation both before and after ~2.2 Myrs. This time is approximately $t_1 + \Delta t_1$ as well as $t_2 - \Delta t_2$ from Table~\ref{tab:gaussfits}, and is thus used as a benchmark for the transition between the two star-forming periods observed in the SFR plots. Before this time, star formation occurs in the same self-similar manner as noted in the baseline and intermediate run. However, the stars formed after this point display a notably different character both from the earlier stars from the same simulation and from those of the other simulations. While essentially no new high mass stars form after this interval, with no new stars above $15 M_{\odot}$ and a $50~\%$ increase in stars between 7 -- 10~\msun{} over 6 Myrs. Over the same period, the population of $1 M_{\odot}$ stars increases by almost a factor of 4, and the population of stars with masses below 0.5~\msun increases by more than an order of magnitude. As a result, the average mass of each star formed before 2.2~Myrs is 9.9~\msun, while the average star mass formed after 2.2 Myrs drops to around 3.3~\msun.

The high magnetic field simulation thus displays a shift in the IMF of stars formed in the later parts of the simulation. Stars formed in this epoch, which aligns with the second peak seen in the SFR, tend to have a lower mass than those formed earlier. The drift observed in the intermediate B-field cloud's average stellar mass towards lower mass (visible in the leftward drift of the vertical dashed lines in the right panel of Fig. \ref{fig:imf1}) may indicate a similar low mass star overabundance in that cloud's second star-forming epoch; however, the two star-forming phases in this run overlap considerably in time, making differentiating the members of the two stellar populations difficult.

As such, the high magnetic field run displays a clear bimodality in the star formation history, along with a corresponding shift in the IMF and average stellar mass. This bimodality is not present in the fiducial run and is seemingly present in the intermediate B-Field run -- though to a smaller degree than in the high B-field run.

\section{Discussion}\label{sec:disc}
As illustrated in the previous section, the presence of stronger magnetic fields significantly altered the geometry, SFE, and IMF of the simulated GMCs, as well as introducing a bi-modality to the SF history. We postulate several processes by which the magnetic fields may produce these effects. 

\subsection{Magnetic Influences on the Scale of the Cloud}\label{ssec:cloudinf}
\begin{figure*}
\includegraphics[width=\columnwidth]{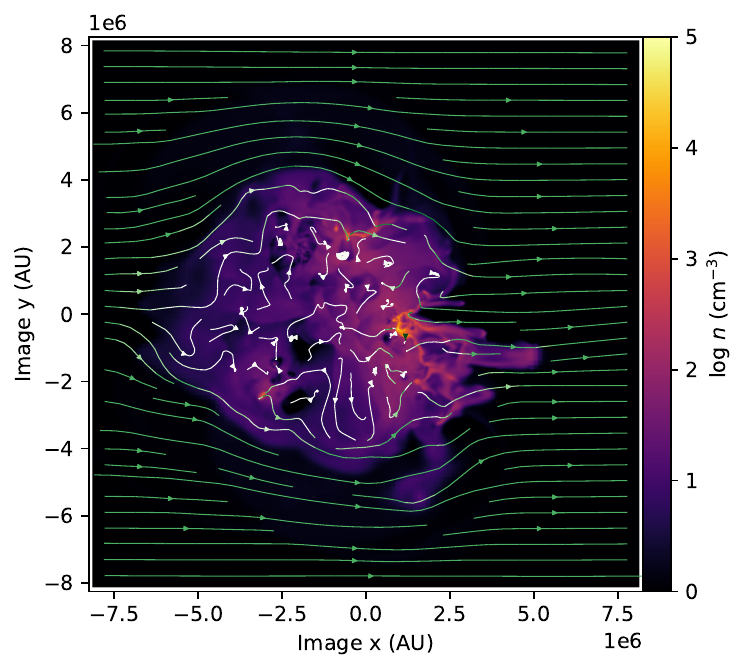}
\includegraphics[width=\columnwidth]{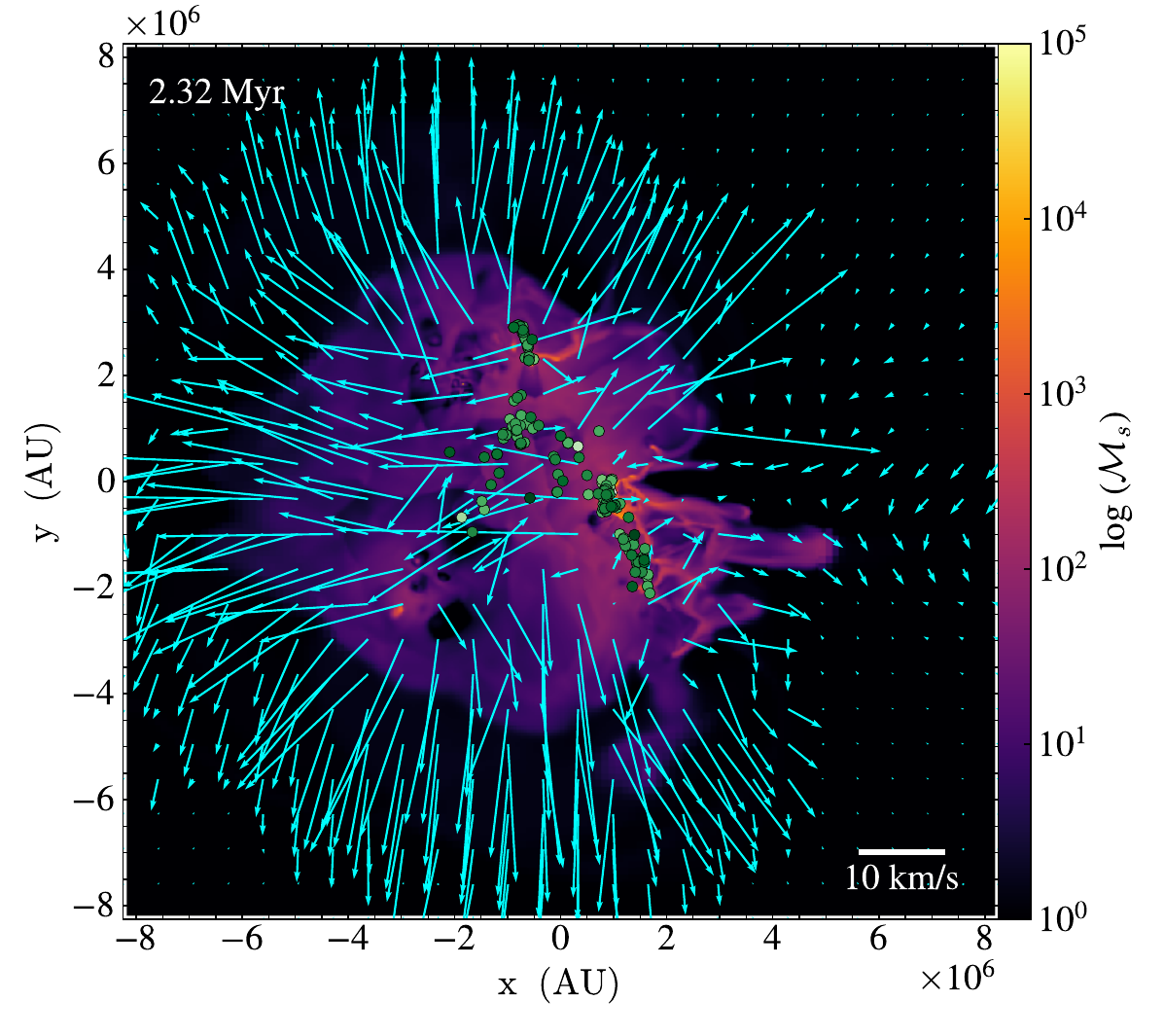}
\caption{A plot of the magnetic field geometry (left panel) and gas velocity field (right panel) along a slice of the fiducial simulation. The magnetic streamlines are coloured according to the magnetic field strength: the white colour corresponds to $1~\mu G$ field strength and dark green corresponds to $100~\mu G$ The medium green field strengths in the low-density gas corresponds to the fiducial background field of approximately $10~\mu G$. Note that the original orientation of the magnetic field in the cloud has been entirely dispersed. The gas ejection appears to be isotropic and not noticeably affected by the magnetic fields.
}
\label{fig:301stream80pc}
\end{figure*}

\begin{figure*}
\includegraphics[width=\columnwidth]{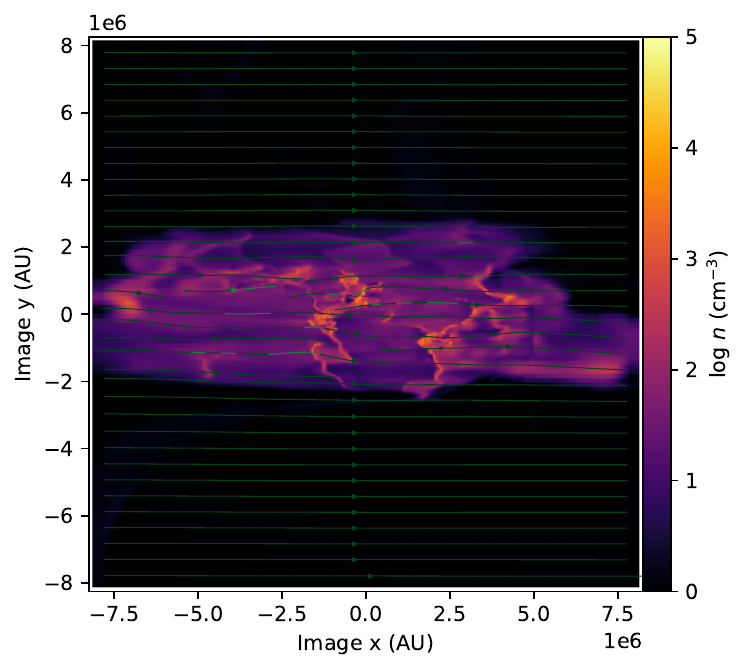}
\includegraphics[width=\columnwidth]{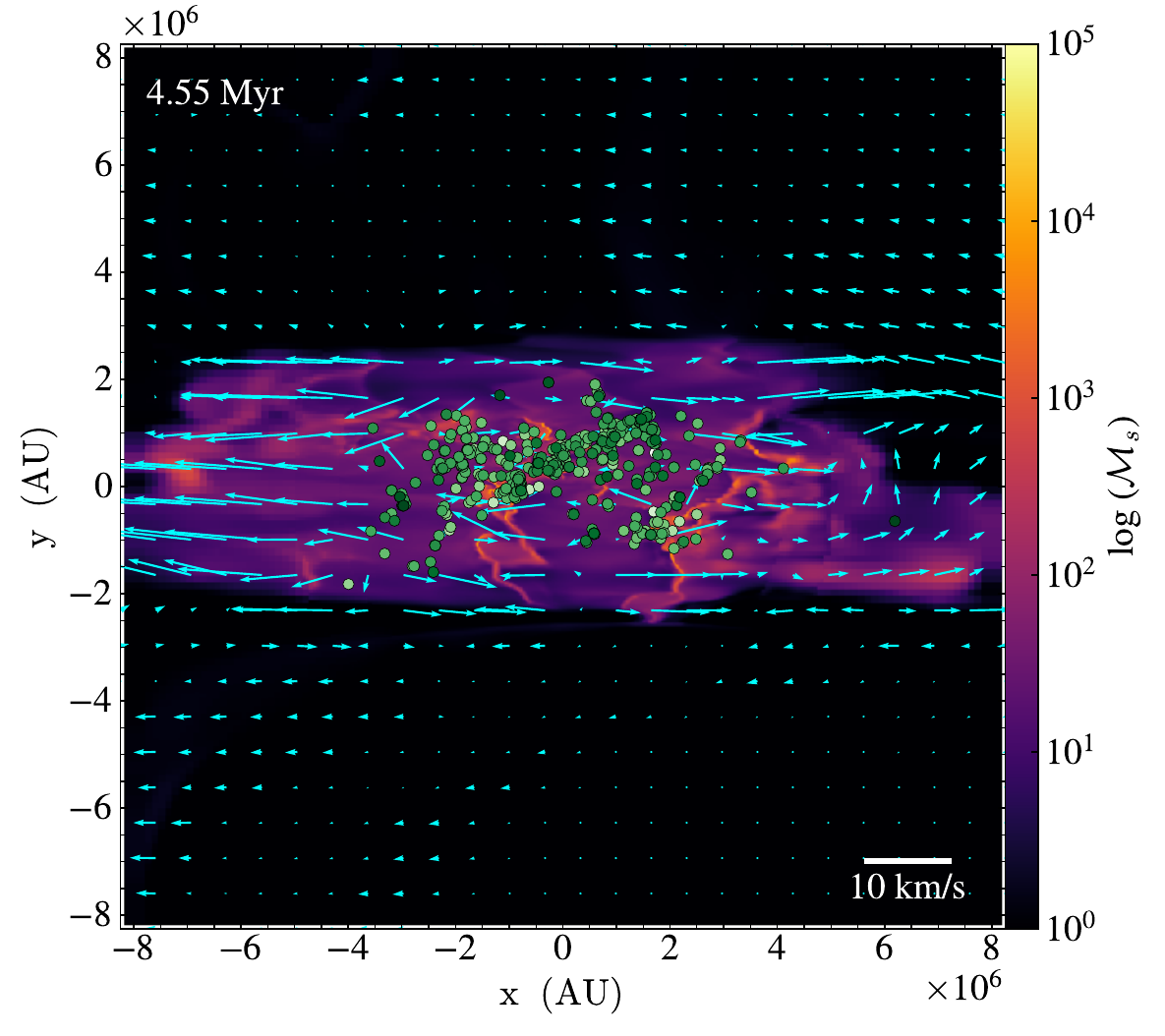}
\caption{A plot of the magnetic field lines and gas velocities along a slice of the cloud in the high B-Field simulation. As in the preceding figure, in the magnetic streamline figure (left), white corresponds to $1\mu G$ field strength and dark green corresponds to $100 \mu G$. Note the intact geometry of the magnetic field and the strongly collimated ejection of gas parallel to the field lines. The right panel also shows the locations of sink particles, coloured by their masses.}
\label{fig:302stream80pc}
\end{figure*}

On the cloud-wide scale, gas expulsion in the high magnetic field runs is noticeably suppressed in the directions perpendicular to the magnetic field axis. Analysis of the gas velocities in the fiducial simulation reveals isotropic ejection driven by UV emissions following the formation of the first stars, presented in Figure \ref{fig:301stream80pc}. Note the strong gas ejection in all directions and the strongly non-uniform magnetic field orientations. In the strong magnetic field run the gas velocity field was entirely aligned with the magnetic field lines, with no significant lateral motion. This can be observed in Fig. \ref{fig:302stream80pc}. It seems evident that the magnetic tension of the strong B-fields effectively resists lateral gas motion, creating a one-dimensional, ``tube-like'' gas ejection mode. We believe this effect to be the source of the bimodality observed in the star formation history of these clouds.

The dramatic collimation of ejected gas in the centre of a strongly confined cloud necessarily results in large bulk flows of ejecting material. In a non-confined cloud, the basic geometry of isotropic expansion quickly reduces the densities of ejected gas to well below the star-forming threshold. The mono-directional bulk outflows of the magnetically confined clouds, in contrast, remain dense for significantly longer, allowing a longer overall star formation period. 
Further, gas in the centre-most region of the clouds can experience additional complications. As star formation continues throughout the elongated cloud, the formation of intermediate and high-mass stars further to the periphery of the cloud is a statistical inevitability. When this occurs along the magnetic axis, strong UV emissions from these stars produce pressure fronts that act to partially ``plug'' the ends of the one-dimensional ejection path. Such behaviour can be observed in Figure \ref{fig:confine}; to the left of the frame, around (x=-18 pc, y=-2 pc) is a UV bubble swept out by several newly formed central stars. The feedback that sculpted this particular bubble is dominated by the emissions of a 56.4~\msun{} and 29.9~\msun{} pair of sink particles. Fragmentation studies anticipate the primary stars formed by these particles to be 22~\msun{} and 12~\msun{} \citep{HeRG:2019}. The gas in the central region of this cloud is trapped between the ``hammer'' of the pressure front in the $-\hat{x}$ direction and the ``anvil'' of the cloud bulk in the  $+\hat{x}$ direction. In a low-field cloud, this gas would likely be ejected laterally; however, in the high-field runs, the magnetic field confinement in the $\pm \hat{y}$ and $\pm \hat{z}$ directions prevents this. The result is a region of gas trapped at star-forming densities for a significant portion of the cloud's evolution. As noted in Section \ref{sec:res}, this result is indeed observed, with dense gas remaining in the higher field clouds for much longer periods than in the fiducial, enabling their notably longer overall star-forming period. 

While the simple prolonged presence of dense gas in the high field clouds justifies their longer overall star formation period, it is insufficient to explain the bi-modality of the observed stellar formation. We believe, however, that this is also a result of magnetic confinement. As stated in Section \ref{sec:res}, the regions of ``trapped'' gas in the higher field runs were characterized by a chaotic web of small-scale filaments. These filaments appear to result from the complex interactions of a large number of UV-driven pressure fronts within the trapped gas. This results in a turbulent velocity structure independent of the turbulent velocity field from the initial condition, which characterizes the initial cloud evolution. As such, the second phase of star formation is fueled by recycled gas which obeys a distribution of velocities controlled entirely by interactions with the magnetic field. This provides a fundamentally different seed environment for the formation of the second population of stars. Compared to the initial turbulent field, the 2nd phase turbulent field varies on much smaller length scales, in turn prompting a larger number of more localized density enhancements, rather than large-scale filaments.  

In such an environment, large contiguous filaments of the type observed earlier in the simulations and in the fiducial run are likely to be disrupted and fragmented by turbulent motion or UV shock fronts before they can collapse sufficiently to form stars. This results in a lower gas supply in the filaments. However, this factor alone does not necessarily produce smaller stars, as smaller filaments could simply fragment into fewer pre-stellar cores, preserving average stellar mass -- assuming these filaments have more total gas mass than the largest possible pre-stellar cores. While this condition is likely true for some of the smallest filaments, it is not universally true, prompting further analysis of magnetic influences.

We note that previous studies have found molecular clouds in regions of high thermal pressure to experience strong confinement and the suppression of HII region expansion into the ambient medium \citep{Iliev:2009, Barnes:2020}. It is natural to consider if the large-scale confinement observed in our simulations is the result of similar influences by high ambient magnetic pressure. However, we do not believe this to be a significant influence in our runs. Our magnetic initial conditions result in magnetic field strengths in the ambient medium which are in equilibrium with those of the outer edge of the cloud; within the cloud, the magnetic field strengths after cloud relaxation are fairly uniform on large scales(See Section ~\S~\ref{ssec:sims}), except for strong enhancements in the local region of dense filaments. As a result, while the magnetic pressure itself is relatively high (comparable to the HII thermal pressure for the highest magnetic field run), the magnetic pressure gradients are weak or nonexistent. As a result, the magnetic pressure is not believed to be a significant source of confinement, and HII regions are observed to expand into the ambient medium without noticeable suppression, even perpendicular to the magnetic field lines. In such a region, with high field strength but minimal magnetic pressure gradient, we believe the magnetic tension is the primary source of cloud confinement.

We note, however, that strong magnetic pressure gradients are observed on smaller scales of the simulation. The initial formation of HII regions often results in the creation of voids in magnetic pressure. While the strong magnetic pressure gradients at the boundaries of these voids likely slow HII region expansion (per \cite{Krumholz:2007}), the HII shock front decouples from the magnetic void fairly rapidly, and thereafter expands without significant magnetic pressure constraint through the remainder of the cloud, owing to the previously mentioned low large-scale magnetic pressure gradients. However, in the regions of these magnetic voids, the magnetic pressure gradients drive turbulent gas flows as the region returns to magnetic pressure equilibrium. We believe these induced flows are likely major contributors to the aforementioned turbulent, chaotic state of the ``trapped'' gas observed in the strongest magnetic field run.

While the magnetic influences discussed in this section are able to explain the differences in large-scale geometry and behaviour of the more strongly magnetized clouds and even provide a source for the prolonged star formation timescales, they are insufficient to explain the differences in the stellar IMF observed in the second star-forming period. To explain this observation, we must also consider the influence of magnetic effects on smaller scales.

\begin{figure*}
\centering
\includegraphics[width=\columnwidth]{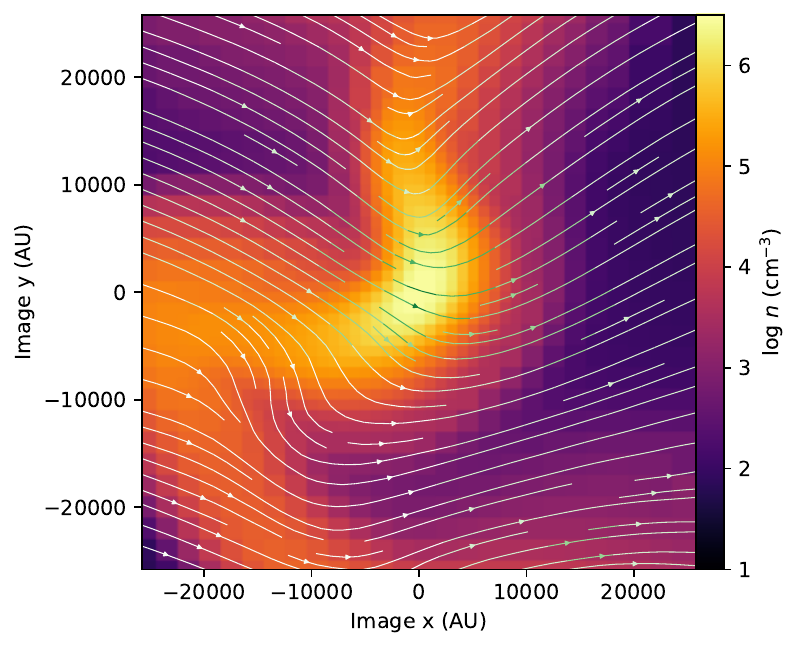}
\includegraphics[width=\columnwidth]{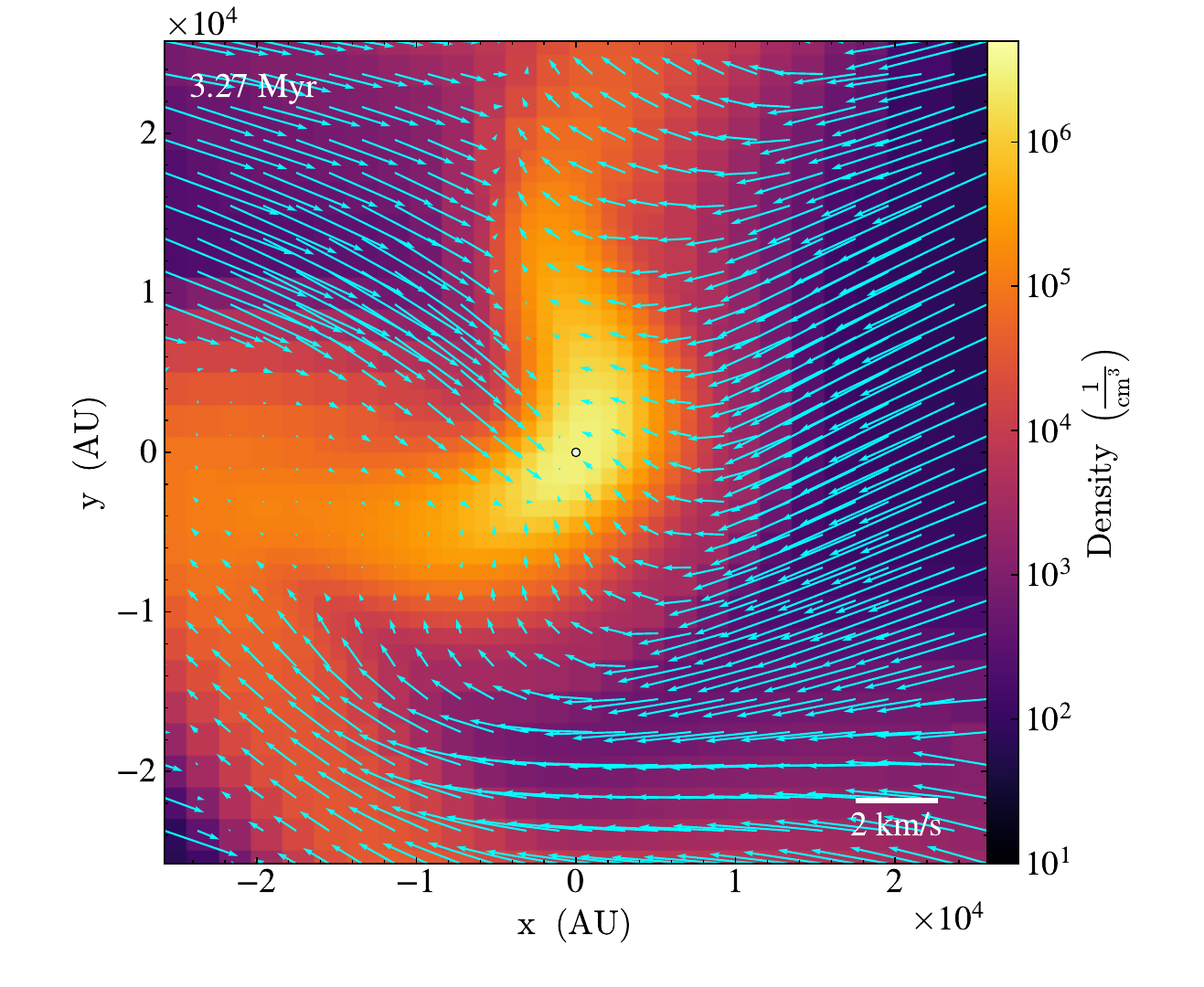}
\caption{A comparison of the magnetic field orientation (left panel) and gas velocity (right panel) in the local region of a newly-formed and accreting sink particle in the second phase of star formation. In the magnetic streamline plot (left), white corresponds to a field strength of $50~\mu G$, dark green is $500~\mu G$.}
\label{fig:stream1pc}
\end{figure*}

\subsection{Magnetic Influences on the Scale of Pre-stellar Cores}\label{ssec:prestellarinf}

The influences of the magnetic field manifest in additional ways on smaller scales. Similarly to their role in confining the cloud on the large scale, strong fields also restrict gas flow on the scales of the star-forming filaments. This effect can be seen in Fig. \ref{fig:stream1pc}, a plot of the magnetic streamlines and velocity field around a newly formed second-epoch star in the high B-field run. The velocity quivers in the right panel show that gas motion is remarkably confined along the magnetic field lines, even when these field lines produce an unusual geometry. Despite the presence of a fairly large untapped gas supply within the filament, the velocity field reveals that the magnetic fields are preventing nearly any of this gas from accreting along the length of the filament onto the newly-formed protostellar core. We note that flux freezing ensures that such filament formation also occurs even in the low-field run, but the stronger fields in the higher B-field simulations result in stronger confinement and less perpendicular inflow. In fact, gas near the top of the filament in Fig. \ref{fig:stream1pc} is observed to be flowing away from the young star. This outflow is not caused by feedback, as the flow is unidirectional and the protostellar core at this timestep is not yet massive enough for any feedback generation. It appears to instead be driven by magnetic pressure from the highly tensioned field lines at the top of the figure. This strong magnetic support, sufficient to suppress and even reverse gas accretion at scales as small as 0.05pc from a newly formed protostellar core, naturally has a commensurately strong influence on the stellar population. The diminished gas flow will limit the final masses achieved by these stars, helping produce the lower average stellar mass observed in the higher B-field simulations. 

The accretion suppression is especially influential in the second star-forming epoch. We support this by comparing local $\mu$ values for the larger filaments of the early phase with the smaller filaments of the second phase. Assuming flux-freezing, the strength of the magnetic field scales with a power of the gas density, $B \propto \rho^\kappa$ \citep{Crutcher:1999}. For regions with relatively uniform field strengths, as is the case within a filament, the magnetic energy is described by ${\cal B} \propto B^2R_{Fil}^3 \propto \rho^{2\kappa}R_{Fil}^3$, where $R_{Fil}$ characterizes the size scale of the filaments. In such a depiction, the binding energy is $W \propto \rho^2 R_{Fil}^5$. As a result, we find the following relation for $\mu$.
\begin{equation}
    \mu = \sqrt{\frac{|W|}{\cal B}} \propto \sqrt{\frac{\rho^2R_{Fil}^5}{\rho^{2\kappa}R_{Fil}^3}} \Rightarrow \mu \propto R_{Fil} \rho^{1-\kappa}.
\end{equation}
In our simulations, the peak density of the filaments is controlled by the sink particle formation criteria, resulting in comparable densities in all cores/filaments, regardless of size. As a result, the effective $\mu$ values of smaller cores in the same simulation are lower. This trend is also found in zooming-in simulations of core formation \citep{Hennebelle2018}. 

This implies that physically smaller cores/filaments in the second phase of star formation, are more strongly supported against gravitational collapse by the magnetic fields and experience greater accretion suppression. As a result, the stronger magnetic fields of the non-fiducial simulations, while present for the entirety of the simulation, decrease the average mass of the second-phase stars much more significantly than the stars of the first SF phase.

\begin{figure*}
\centering
\includegraphics[width=\columnwidth]{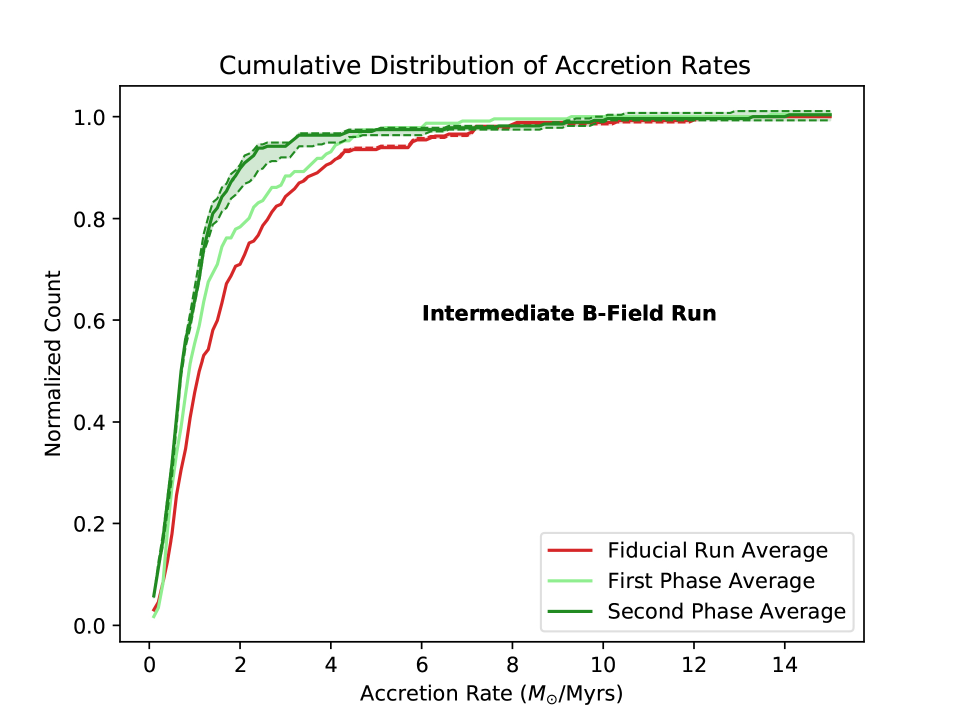}
\includegraphics[width=\columnwidth]{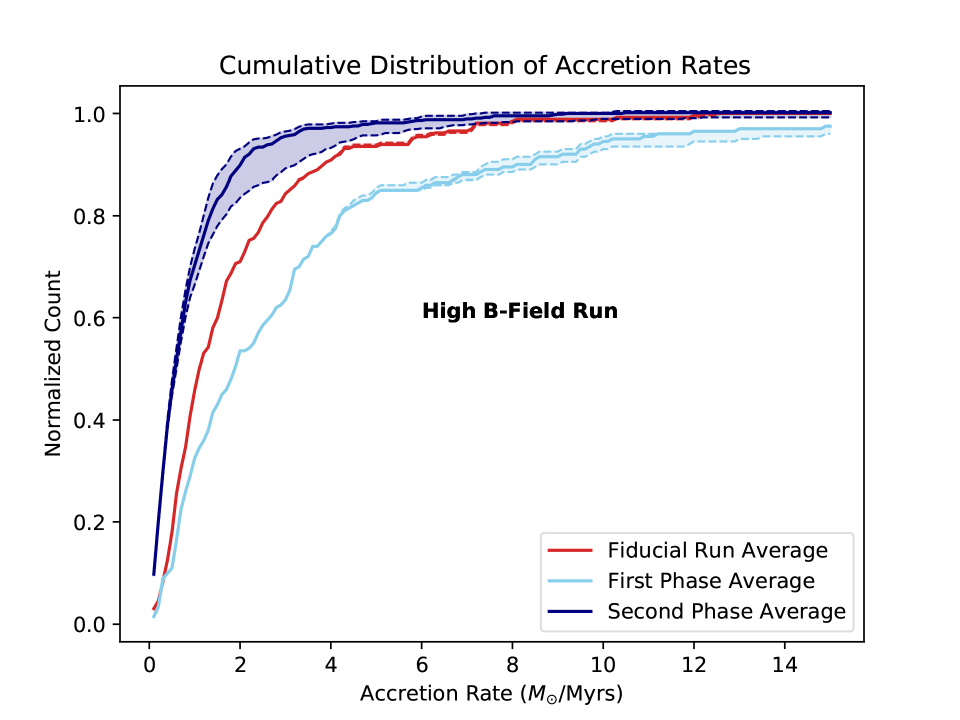}
\caption{The cumulative distribution function of accretion rates (x-axis) onto sink particles for the intermediate- (left panel) and high- (right panel) B-field runs. The corresponding distribution for the fiducial run is included (red solid line) in both plots for reference. Discrete simulation output rates introduce uncertainty in the accretion rates resulting from imprecision in the sink formation time. Dashed lines represent 75\% confidence intervals on these accretion rates. Note the systematically lower accretion rates onto the sink particles during the second phase of star formation in both simulations.}
\label{fig:accreterate}
\end{figure*}

Bulk analysis of the stellar populations provides further evidence of accretion suppression in the stronger B-field runs. Fig. \ref{fig:accreterate} shows the cumulative distribution function of the mass accretion rate for stars (sinks) formed during the first and second star-forming epochs of the intermediate and high B-Field run, plotted in relation to the accretion rate CDF of the fiducial run. We observe that the CDF for the second phase peaks much more rapidly, indicating that the typical average accretion rate of this second stellar population is significantly lower. Conversely, the initial phase of the intermediate run displays accretion rates consistent with fiducial results. Accretion in the first phase of the high B-field run is similarly not suppressed and, instead, displays higher accretion rates than in the fiducial case. There are several factors that could cause this increased accretion, primarily the effects of magnetic braking, which causes increased gas radial infall by suppressing disk formation \citep{Mestel1956}. However, further study is required to provide a robust understanding of this effect. Overall, these observations provide robust evidence that the strong magnetic fields greatly suppress gas accretion for stars formed in the second star-forming epoch. 

It is further worth noting that this confinement would be expected to result in a smaller ``effective accretion radius'', that is, in a strong field environment, a star can only effectively accrete material from a much smaller region than an equivalent mass star could in a fiducial environment. This means a filament of the same mass and density would be expected to fragment into a greater number of stars in a strong field environment. This is consistent with the larger number of small stars formed in the high field runs, although the difficulty in finding directly comparable filaments in the different simulations renders a detailed quantification of this effect unfeasible. 

We note that previous observational studies of molecular clouds presented in \cite{Palau:2021} find that, on the scale of 1000-4000 AU, smaller, lower mass filaments and prestellar cores tend to dominate in more strongly magnetized regimes, with larger cores and filaments dominating in weaker field regimes. As such, the smaller size of the second-phase filaments (which have more dominant magnetic influences) in these simulations is consistent with the observed filaments in the referenced work.

We must also consider the potential influence of unresolved fragmentation of the pre-stellar cores. As noted in ~\S\ref{ssec:imf}, our 1000~AU resolution limit is insufficient to resolve the formation of individual stars, which occurs on scales of the order of a few hundred AU \citep{Krumholz2016}. Previous work in \cite{HeRG:2019} has indicated that scaling the core mass by 0.4 reproduces an IMF consistent with a properly normalized Salpeter high-mass end. However, the simulations in this work all possess relatively weaker magnetic fields, on par with those of our fiducial simulation. It is very possible that the cores in the more strongly magnetized clouds undergo different fragmentation, which would alter the true IMF. However, existing high-resolution studies by \cite{Cunningham:2018} suggest that fragmentation in dense regions is insensitive to the magnetic field on scales below 0.5~pc. Additionally, ``zoom-in'' simulations by \cite{Heinprep} have aimed to probe prestellar core fragmentation based on similar simulations as in this work. These simulations have drawn pre-stellar core regions from large-box simulations (including some drawn from the runs presented in this work), and simulated the formation of individual stars with resolutions down to 7~AU; Detailed discussion of the methods and conclusions of these studies can be found in their respective papers. The authors suggest that strong magnetic fields on the scale of the pre-stellar cores cause increased fragmentation. According to their result, the smaller cores of the second phase of star formation, which are more strongly influenced by the magnetic fields, would be expected to experience more fragmentation, rather than reduced fragmentation which would be necessary to reconcile them with the IMF of the first phase. As a result, while unknown differences in sub-grid fragmentation could alter the scaling between the core mass function and the IMF, current literature suggests that applying the same scaling relation for clouds of all magnetic field strengths is not unreasonable, and may likely be an underestimation of magnetic influence. Nevertheless, further higher-resolution analysis remains a prominent avenue for future work.

\subsection{Comparisons with Previous Studies}\label{ssec:compare}

Overall, the effects on the SFE and the bimodality of the stellar population in the high B-field clouds seem to be the result of magnetic effects on both cloud-wide and localized scales. On the bulk scale, the confinement of the cloud traps significant quantities of star-forming gas that would otherwise be ejected by feedback. This enables a longer star formation period, increasing the SFE. Feedback from the first population of stars and interactions with the magnetic fields create a turbulent environment composed of many small filaments. On the local scales, magnetic support efficiently suppresses gas accretion onto newly formed stars, causing lower final stellar masses. This influence is especially prominent in the later stages of cloud evolution, where the magnetic fields exercise a more dominant influence on fragmentation. 

In comparison to previous work, particularly \cite{Kim:2021}, we find many points of agreement in our simulation suite. In both works, the star formation timescales increase for the simulations with the strongest magnetic fields. In addition, star-forming filaments are observed to orient predominantly perpendicularly to the magnetic fields. Further, gas ejection in both works is observed to occur preferentially along the magnetic field lines. These similarities, present in simulations with significantly different initial cloud parameters and computational methods are an encouraging sign for the validity of the simulations. However, there are also some differences that bear mentioning.

While the simulations presented in this work do not produce the reduced total SFE found in \cite{Kim:2021} and instead yield equal or increased values, these results are not as inconsistent as they may appear. If we limit our analysis to the first phase of star formation, the phase shared by all the simulations, we observe a decrease in the SFE for the higher B-field runs. Quantification of the magnitude of this first-phase-only SFE is difficult, as disentangling the 2 fairly co-temporal phases of star formation in the intermediate B-field run. However, it nevertheless seems evident that, if the 2nd phase of star formation were discounted, our results would be consistent with \cite{Kim:2021}. This could indicate that the production of the second phase is sensitive to differences either between our initial conditions or computational methods. 

One such possibility is the differences in turbulent support, as in this work the cloud is initially roughly in virial equilibrium: $\alpha_{0} \equiv K/|W| = 0.4$, whereas in \cite{Kim:2021}, $\alpha_{0} = 1$ \footnote{Note that these two values are reported using our formulation of $\alpha$ = $K/|W|$, whereas the values of $\alpha$ in \cite{Kim:2021} are reported as $2K/W$.}. The stronger turbulent support in that work may serve to produce a cloud that is less bound by gravity\footnote{Once we consider the magnetic field contribution in addition to turbulence, in our work, we have $\alpha_{tot} = (K+{\cal B})/|W| \sim 1.28>\alpha_0$.}, hence is less able to sustain star formation during the second phase as the gas disperses more rapidly. Differences in UV feedback modeling may also play a role. The stellar feedback recipe in \cite{Kim:2021} is less crude than what it is utilized in our work, including realistic emission of H$_2$ photodissociating radiation in the Lyman-Werner bands in addition to Lyman continuum photo-ionizing photons as a function of sink mass. Our method assigns a UV ionizing luminosity to each sink particle according to a single power law as a function of sink mass, somewhat overestimating the UV feedback from low-mass stars (to account for missing feedback from low-mass stars, for instance from protostellar outflows, see \S~\ref{sec:sim}). 
While these two feedback recipes appear to yield comparable results in the fiducial cases (where emissions from massive stars dominate), it is possible that the differences in low-mass stellar feedback have significant effects during the second phase of SF in which stars are systematically lower in mass. Additionally, work by \cite{Guszejnov:2021} finds that pre-stellar jets play an important role in regulating star formation and reproducing the empirical IMF. While our feedback recipe attempts to account for missing feedback from these outflows by slightly overestimating UV feedback from low mass starts, more rigorous modeling of this feedback source may impact our results.

Finally, we point out that the two studies use different codes and numerical methods, leading to significant differences in grid resolution. The simulations presented in \cite{Kim:2021} use a fixed grid simulation with a grid width of $\Delta x = 0.31~{\rm pc}$, while we use an AMR grid, resolving the Jeans length with 10 cells and a resolution at the most refined level of $\Delta x_{min} =1000~{\rm AU}$ (0.0048~pc). 
We estimate that the width of the large filaments during the early phase of SF or in weakly magnetized clouds is about 0.5-1~pc, while during the second phase of SF, in the strongly magnetized case, they are much thinner, with widths below 0.05~pc being the norm. Hence, high resolution is particularly important to resolve star formation in such turbulent and highly magnetized environments as evidenced by comparing the left and right panels in Fig.~\ref{fig:confine} and by Fig.~\ref{fig:frames1}. We note, however, that the high-resolution requirements of this finely structured environment result in large increases in computational requirements. As listed in Section \ref{sec:binding}, the strong B-field run required more than 6 times the CPU hours of the fiducial simulation, complicating the study of strongly magnetized clouds. 

This effect, namely that a stronger magnetic field increases the SFE, is also found in GMCs that are formed out of converging flows in the diffuse warm atomic medium \citep{Zamora-Aviles2018}. The authors find that stronger fields reduce the turbulence generated by the instabilities in the compressed layer, thus expediting SF activity. While we believe the increased SFE in our simulations results predominately from other influences of the magnetic field, these findings are nonetheless interesting additional examples of magnetically induced SF enhancement.

Overall, the presence of a prominent bi-modality and enhanced SFE, as observed in our simulations, could be produced by a variety of factors. Further simulations with different initial conditions, particularly cloud virial parameters, UV feedback recipe, and more resolution studies, are required to determine which variables are crucial to capture the presence of a second phase of star formation.

\section{Summary and Conclusions}\label{sec:sum}

We have conducted AMR radiation-MHD simulations of the collapse of a suite of GMCs varying the initial magnetic field strength from $\mu\sim 1$ to $\mu \sim 5$, within the range observed in local star-forming molecular clouds. 
We model ionizing UV radiation from individual massive stars that self-consistently form from the filamentary collapse of the clouds. Energetic processes from these stars drive outﬂows in the gas around them and quench star formation.

We find that the effect of enhancing the magnetic field is twofold. Firstly, the cloud expansion due to UV feedback from massive stars is confined in the direction perpendicular to the magnetic field, producing a quasi-cylindrical cloud geometry, aligned with the magnetic field. As shown in \cite{HeRG:2019}, UV feedback from massive stars determines the duration of star formation in the cloud, roughly proportional to the cloud size. The magnetic confinement observed in these simulations reduces gas expulsion, thereby retaining dense gas for a longer period. 
This results in progressively prolonged periods of star formation in clouds with higher B-field strengths. Further, the confined gas is subjected to the interactions of a large number of UV bubbles with the magnetic fields, creating turbulence and fragmentation patterns characteristically different from the initial conditions.

Secondly, a clearly bimodal SFR is observed in the cloud with the strongest magnetic field, and a similar, though less prominent, feature is observed in the intermediate B-field simulation. The stars formed in this second phase of star formation obey a different IMF, with a significantly lower average mass. It appears that the same mechanism of gas confinement works at smaller scales, reducing the growth rate of sink particles and thereby reducing the mean mass of the stars. This is especially evident in the second burst of star formation, where gas fragmentation is controlled by a density and velocity structure primarily created by the interaction between the magnetic field and the energy injection from massive stars, producing hot photoionized bubbles and winds. The smaller mass cores/filaments observed in this second phase are more strongly affected by the magnetic field, that roughly remains of similar strength during the cloud evolution, causing the gas accretion onto protostellar cores during the second phase of SF to be more strongly suppressed. 
This has an interesting effect on the evolution of the overall IMF of the cloud. Initially, the IMF evolves with a Salpeter slope at the high-mass end: both high-mass and low-mass stars are formed, as found for the cases with a fiducial magnetic field. However, during the second episode of star formation, only low-mass and intermediate-mass stars are formed. A Salpeter IMF is, nevertheless, maintained because the high-mass end of the IMF does not evolve significantly.
Higher resolution simulations and/or zoom-in simulations on sink particles in the strong magnetic field cases are required to conclude whether the shape of IMF at low masses is universal or depends on the magnetic field strength. 
In \cite{HeRG:2019}, for the weaker magnetic field case, we argued that the fragmentation of sink particles can reproduce a Kroupa IMF at low masses. However, in the stronger magnetic field cases presented in this work, during the second phase of SF we form significantly more low-mass cores than in the fiducial run. These cores will presumably form low-mass stars, but their fragmentation can also differ significantly from the case with low B-field. 

A final noteworthy speculation regards the puzzling origin of multiple stellar populations in GCs \citep{Bastain:2018}.
The observed bi-modality of SF in the strong magnetic field case and the bi-modal formation of low-mass stars, mostly produced during the second peak of SF, could be ingredients not previously considered important for the solution of this long-standing mystery. Chemical enrichment of the gas that forms the second stellar population by winds and ejecta (though not supernovae) from the first population could help explain the multiple chemical abundance populations observed in GMCs. To investigate this potential link, more work is necessary to model possible chemical enrichment differences/patterns between the first and the second populations of low-mass stars in our simulations. We emphasize that there are many other proposed mechanisms that could help explain this puzzle \citep{Bastian2015,Renzini:2015,Kroupa:2018,Wang:2020}. Further, as is the case in almost all previous studies, the results presented here are based on simulations with somewhat idealized initial conditions. It remains to be seen if bi-modal star formation remains present in simulations with increasingly realistic initial conditions. Nonetheless, the mere existence of a bi-modality of SF is an intriguing starting point for further investigation.

\section*{Acknowledgements}
We thank Dr. Sam Geen for sharing his version of the RAMSES code. CCH acknowledges the support by the NASA FINESST grant 80NSSC21K1850. CCH and MR acknowledge the support of the NASA grant 80NSSC18K0527. The authors acknowledge the University of Maryland supercomputing resources (http://hpcc.umd.edu) made available for conducting the research reported in this paper.

We thank the developers of NumPy \citep{vanderWalt2011}, Matplotlib \citep{4160265}, and YT \citep{Turk2011} for their extremely useful open-source software.

\section*{Data Availability}
The data underlying this article were accessed from the University of Maryland supercomputing resources (http://hpcc.umd.edu). The derived data generated in this research will be shared upon reasonable request to the corresponding author. The software used to do the analysis in this paper is \textsc{ramtools}, a toolkit to post-process \ramses{} simulations based on the YT toolkit (\url{https://ytproject.org/doc/index.html}) and is available to download from \url{https://chongchonghe.github.io/ramtools-pages/}.

\bibliographystyle{mnras}
\bibliography{reference,BIB_HE,he2}

\appendix

\section{Cloud Parameters and Relaxation}
\label{sec:binding}

Table~\ref{tab:allparam} shows the initial conditions for the clouds in the different runs presented in this paper. The parameters with subscript '0' refer to the initial conditions ($t=0$) with non-singular isothermal spheres, while the subscript '1' refers to the corresponding values calculated at the end of the relaxation period ($t=t_{relax}$), just before the beginning of star formation. During this relaxation period, the strength of gravity is reduced, allowing turbulence to develop before the cloud's collapse. This process produces a less artificially uniform initial condition. However, the expansion of the cloud during this phase affects the bulk density. In the fiducial simulation presented in this work, the lack of magnetic confinement allows greater cloud expansion during the relaxation phase, reducing the densities more strongly than in the magnetized clouds. To counteract this, the relaxation timescale for the fiducial run is reduced such that the cloud density at the beginning of star formation is roughly consistent with the strongly magnetized clouds. Note the inclusion of the ``M-C'' cloud which represents the fiducial cloud in perfect correspondence to that presented in \cite{HeRG:2019}. In this work, the fiducial cloud was relaxed for only half as long as the M-C cloud, so that the mean density at the end of the relaxation period was closer to those of the Intermediate and High B-Field runs. This different relaxation timescale appears to have minimal impact on the virial state of the cloud at the end of relaxation.

\begin{table*}
\caption{\label{tab:allparam}
The initial conditions of the simulated GMCs.
The clouds has initial mass at $t=0$ of 43,000 \msun{} and average density of around 300 \pcc{}, radius of 10 pc, $\alpha_0 = 0.35$. Note that $\alpha_{tot} = (K+{\cal B})/|W|$, including both turbulent and magnetic support, is a better measure of support against gravity for clouds where magnetization is significant. The parameters with subscript `0' refer to the initial conditions, while the subscript `1' refers to the corresponding values calculated at the end of the relaxation period, just before the beginning of star formation.
}
\begin{tabular}{lllllllllll}
 job name & $t_{relax}$ (Myrs) & $\alpha_{tot,0}$& $B_{0}$ ($\mu$G) &  $\mu_0\hide{ = \sqrt{|W|/{\cal B}}}$ & $n_1$ ($\pcc{}$) & $R_1$ (pc) & $\alpha_{tot,1}$ & $B_1$ ($\mu$G) & $\mu_1$ & CPU hours \\
\hline
     Fiducial B-Field & 2.11 & 0.39  & 11.7  &  5.18 & 152 & 12.5 & 0.27 & 10.6 & 3.69 & 72,000  \\
     Intermediate B-Field & 4.22 & 0.50  & 23.4 &  2.59 & 127 & 13.27 & 0.44 & 21.0 & 1.77 &  96,000     \\
    High B-Field & 4.22 & 1.28  &  58.4 &  1.04 &  141 & 12.8 & 1.18 & 53.6 & 1.05 &  460,000  \\
    M-C \citep{HeRG:2019} & 4.22 & 0.39 &  11.7   &  5.18 & 95 & 14.6 & 0.23 & 8.9 & 3.4 & -- \\
\end{tabular}

\end{table*}

\label{lastpage}
\end{document}